\documentclass[aps,nofootinbib,superscriptaddress, showpacs,preprintnumbers,  twocolumn, nofootinbibt]{revtex4-2}
\usepackage{eurosym}
\usepackage{amsmath}
\usepackage{bm}
\usepackage{amsfonts}
\usepackage{amssymb}
\usepackage{float}
\usepackage{graphicx}
\usepackage{subfig}
\usepackage{caption}
\usepackage[export]{adjustbox}
\usepackage[hyperindex,breaklinks]{hyperref}
\usepackage{eurosym}
\usepackage{hyperref}\hypersetup{
    colorlinks = true,
    linkcolor = blue,
    anchorcolor = blue,
    citecolor = blue,
    filecolor = blue,
    urlcolor = blue,
    breaklinks=true
   }
\graphicspath{ {folder} }

\def\be{\begin{equation}}
\def\ee{\end{equation}}
\def\bea{\begin{eqnarray}}
\def\eea{\end{eqnarray}}

\begin{document}

\title{Warm inflation in a Universe with a Weylian boundary}
\author{Teodora M. Matei}
\email{teodora.maria.matei@stud.ubbcluj.ro}
\affiliation{Department of Physics, Babes-Bolyai University, Kogalniceanu Street,
		Cluj-Napoca 400084, Romania,}
\author{Tiberiu Harko}
\email{tiberiu.harko@aira.astro.ro}
\affiliation{Department of Physics, Babes-Bolyai University, Kogalniceanu Street,
		Cluj-Napoca 400084, Romania,}
\affiliation{Astronomical Observatory, 19 Ciresilor Street,
		Cluj-Napoca 400487, Romania}

\date{\today }

\begin{abstract}
We investigate the influence of boundary terms in the warm inflationary scenario,
by considering that in the Einstein-Hilbert action the boundary can be described in terms of a Weyl-type geometry. The gravitational action, as well as the field equations, are thus extended to include new  geometrical terms, coming from the non-metric nature of the boundary, and depending on the Weyl vector, and its covariant derivatives. We investigate the effects of these new boundary terms by considering the warm inflationary scenario of the early evolution of the Universe, in the presence of a scalar field.
We obtain the generalized Friedmann  equations in the Universe with a Weylian boundary by considering the Friedmann-Lemaitre-Robertson-Walker metric. We consider the simultaneous decay of the scalar field, and of the creation of radiation, by appropriately splitting the general conservation equation through the introduction of the dissipation coefficient, which can depend on both the scalar field, and the Weyl vector. We consider three distinct warm inflationary models, in which the dissipation coefficients are chosen as different functions of the scalar field and of the Weyl vector.     The numerical solutions of the cosmological evolution equations show that the radiation is created during the very early phases of expansion, and, after the radiation reaches its maximum value, the transition from an accelerating inflationary phase to a decelerating one takes place. Moreover, it turns out that the
the Weyl vector, describing the boundary effects on the cosmological evolution,  plays a significant role during the process of radiation creation.

\end{abstract}

\maketitle

{
  \hypersetup{linkcolor=blue}
  \tableofcontents
}

\section{Introduction}

Inflationary cosmology emerged as a theoretical framework to address various challenges within the Standard Big Bang model of the early Universe, including the horizon, flatness, and monopole problems.  A. Guth proposed that the Universe underwent a period of rapid exponential growth, driven by the false vacuum decay of a scalar field, called inflaton \cite{G}. Guth's inflationary theory offered a refined perspective on the early Universe, and it improved Starobinsky's de Sitter era \cite{S} in the early Universe cosmology.

The "old inflation" model, which relied on the concept of a false vacuum state, faced challenges concerning the graceful exit problem. Addressing this issue, A. Linde proposed the theory of "new inflation" in \cite{L-1982}, introducing a slow and gradual evolution of the inflaton field, in contrast to the quick decay from a false vacuum state. This slow-roll behavior allows the inflationary phase to be sustained over an extended period, resulting in a smoother exit from inflation, and a more graceful transition to the hot Big Bang phase.
Following the slow-roll inflation concept, A. Albrecht and P. J. Steinhardt introduced the notion of a "flat" scalar field potential in their work \cite{AS}. Nevertheless, A. Linde extended this idea into what is now known as "chaotic inflation" \cite{L-1983}. In this model, the inflaton field's potential energy is governed by a flat potential, and the mass of the inflaton can be relatively large. Further works considering the development of chaotic inflation lead to various inflationary theories, such as hybrid inflation, natural inflation, axion inflation, to name only a few of them \cite{L-1986,L-1994,K,L-1991,L-1993,S-1994,C,F}. For reviews of various aspects of the inflationary theory see \cite{rev1, rev2, rev3,rev4,rev5,rev6, rev7}.

The transition from the inflationary phase to the Big Bang phase involves the transfer of energy from the scalar field to the Standard Model particles. This process, referred to as reheating, and discussed initially in \cite{Alb}, has been extensively studied in the literature \cite{DL,KL,Lyth,H2008, A, M, N1r,N2r,N3r,N4r,N5r, N6r,N7r,N8r,N9r,N10r,N11r,N12r,N13r,N14r,N15r}. During inflation, quantum fluctuations in the inflaton field were stretched to cosmic scales, and therefore became the seeds for the large-scale structures we can observe in the present-day Universe. The theory of inflation has been supported by various observational evidences, such as the precise measurements of the Cosmic Microwave Background Radiation. One can test the predictions of the theory of the inflation by using the current cosmological and astronomical observations, like the results on the CMB temperature anisotropies obtained by the PLANCK collaboration \cite{Pl1,Pl2}, and with BICEP2/Keck-Array data \cite{Bi1,Bi2}. The spectral index of the scalar perturbations is obtained by the Planck collaboration \cite{Pl2} as $n_s=0.9649\pm 0.0042$ at 68\% CL, and it shows no evidence for a scale dependence. Moreover, spatial flatness is confirmed at a precision of 0.4\% at 95\% CL by also taking into account the BAO data. The Planck 95\% CL upper limit on the tensor-to-scalar ratio is $r_{0.002}<0.10$. By combining the Planck data with the BICEP2/Keck Array BK15 data one obtains the constraint  $r_{0.002}<0.056$. These constraints are obtained in the framework of single-field inflationary models within standard Einstein gravity.

 In the context of inflationary theories, the "warm inflation" scenario presents itself as a successful alternative to the standard inflationary and reheating models, by integrating, in addition to the dynamics of the scalar field, the effects of thermal interactions during the process of accelerated expansion. The presence of thermal radiation allows for the energy transfer between the inflaton and the other fields, maintaining a state of thermal equilibrium. Therefore, the transition between the inflationary phase and the radiation-dominated era is possible without a sudden reheating mechanism, required in the "cold inflation" models. The energy dissipation can lead to a gradual decrease in the energy density of the inflaton field, allowing it to shift smoothly to a state where it is no longer driving the accelerated expansion. The theory was first proposed by A. Berera and L.-Z. Fang in \cite{BF}, as they showed that the interaction between the scalar field and a thermal bath of particles could not only sustain inflation, but also explain the observed fluctuations in the CMB radiation \cite{B,H}. This area of research flourished under the contribution of \cite{O, Moss, Zhang, W1,W2,W3,W4,W5,W6,W7,W8,W9,W10,W11,W12,W13,W14,W15,W16,W17,W18,W19,W20,W21,W22,W23,W24,W25,W26,W27,W28,W29,W30,W31,W32,W32,W33,W34,W35,W36,W37,W37a,W38,W39,W40,W41,W42,W43,W44,W45,W46,W47,W48,W49}.
For the past, present and future status of warm inflation see \cite{W48}. Recent developments in the field are presented in \cite{W49}.

The boundary problem of general relativity arises as a need to ensure that the initial conditions in the evolution of the early Universe, and generally of gravitational systems,  are well defined. In the work of G. W. Gibbons, S. W. Hawking and J. W. York \cite{GH, Y}, an additional term was introduced in the Einstein-Hilbert action, which would relate the boundary and the external curvature.

Later on, J. S. Ridao and M. Bellini \cite{BR1, BR2} bring an alternate proposal for the "Gibbons-Hawking-York" term, as they suggest that the boundary expression should be taken into account as a physical source of geometric nature. The influence of boundary terms in the preinflationary Universe is considered in a Weyl-type geometry. In their work, the variation of the Ricci curvature tensor is defined as $\delta R_{\alpha \beta}\equiv \nabla_{\mu} W^{\mu}=\phi(x^{\alpha})$, where $W$ is a tetra-vector of geometric kind, and $\phi$ is an arbitrary scalar field, which accounts for back-reaction effects due to boundary contribution. The field generated by the tetra-vector is gauge invariant under the transformation $\delta \tilde W_\alpha = \delta W_\alpha - \Lambda g_{\alpha \beta }$. Hence, the Einstein tensor is defined in terms of a boundary parameter $\Lambda$, as $\tilde G_{\alpha \beta}=G_{\alpha \beta }-\Lambda g_{\alpha \beta}$ and consequently, the field equations $G_{\alpha \beta }+\Lambda g_{\alpha \beta}=\kappa T_{\alpha \beta}$ are obtained. The role of the boundary term was investigated for the case of an FLRW metric in \cite{B3}, where a barotropic fluid was considered, in order to determine the origin of the cosmological boundary flux parameter $\Lambda $.

The role of the boundary terms in the presence of non-metricity was considered in the framework of $f(Q)$ gravity in \cite{DLS}, leading to the formulation of the $f(Q,C)$ theory, where $Q$ is the non-metricity, and the boundary term $C$ is the difference from the standard Levi-Civita Ricci scalar  $\tilde{R}$. An effective dark-energy sector of geometrical origin can be obtained within the framework of this model, together with an effective interaction between matter and dark energy. One can also reobtain the  thermal history of the universe,  including the matter and dark-energy dominated epochs. The role of the boundary terms in $f(Q,B)$ gravity was further investigated in \cite{CFF}, where by using a variational principle, the field equations were derived, and compared with those of $f(Q)$ gravity in the limit of $B\rightarrow 0$. Interestingly, the  $f(Q,B)=f(Q-B)$ models are dynamically equivalent to $f(R)$ gravity, a situation similar to the case of teleparallel $f(\tilde{B}-T)$ gravity, where $\tilde{B}\neq B$. Furthermore, conservation laws were also derived.  The geometrodynamical effects of considering the boundary term in $f(Q)$ gravity were investigated in \cite{Pal}. For the connection in the coincidence gauge, it was found that the field equations are of fourth-order. Moreover, the fluid components introduced by the boundary can be described  by a scalar field. The cosmological field equations are equivalent to those of teleparallelism with a boundary term in the coincidence gauge. For the connection defined in the non-coincidence gauge, the geometrodynamical fluid consists of three scalar fields.  The dynamical system analysis of the cosmological models framed in the extended teleparallel gravity in the presence of a boundary term, the $f(T,B)$ gravity was considered in \cite{Kad}.  The critical points are obtained for two forms of $f(T,B)$, the first given by the logarithmic form of the boundary term $B$, and the other one is the non-linear form of the boundary term.

The goal of the present investigation is to extend the work of J. S. Ridao and M. Bellini to a warm inflationary scenario by effectively calculating the variation of the Ricci tensor in Weyl geometry, which was postulated in \cite{BR1}. In order to obtain a general description of the influence of the boundary terms, we will use an approach in which we assume a Weylian geometric structure of the early Universe. A spontaneously broken gauge scale symmetry is considered, which leaves the Einstein-Hilbert action invariant under conformal transformations and therefore leads to a non-metric geometry, where the covariant derivative of the metric tensor does not vanish, $\nabla_{\mu}g_{\alpha \beta} \neq 0$. In Weyl geometry, non-metricity emerges naturally as a result of the dynamics of the Weyl gauge field, $\omega_{\mu}$.

Consequently, as a first step in our analysis, we will formulate an extension of the action principle in which the boundary terms are assumed to be described by the Weyl geometry. Then we will consider this theoretical model, and the resulting equations of motion to construct a warm inflationary model for a Universe in which the creation of matter (radiation) takes place due to the decay of the scalar field,  and of the dynamics of the Weyl vector. By introducing a set of dimensionless variables, we explore the cosmological implications given by different mathematical forms of the scalar potential $V(\phi)$. We develop three models which can be solved numerically, and which provide the evolution of the radiation fluid, and of the cosmological expansion of the very early Universe.

The present paper is organized as follows. The derivation of the gravitational field equations in Weyl geometry is presented in Section \ref{field}. The general equations which describe the creation of the radiation component are derived in Section \ref{action}. Additionally, numerical solutions are obtained in Section \ref{models} for three different models, which show similar behaviours for the functions that describe our system.  The conclusions regarding our results and further discussions, are considered in Section \ref{conclusion}.

\section{Weyl geometry, gravitational action, and boundary terms} \label{field}

In the present Section we concisely review the fundamentals of Weyl geometry, of the Weyl geometric gravity, and of the Hilbert-Einstein variational principle, which we will use in developing the geometrodynamical approach necessary for the description of the effects of the boundary term on the gravitational and cosmological dynamics. We also point out the possibility of obtaining scalar fields of geometric origin coming from Weyl geometry. We use the Landau-Lifshitz \cite{LaLi} metric conventions, and definitions of the physical and geometrical quantities.

\subsection{Brief introduction to Weyl geometry}

\paragraph{Conformal transformations.} Conformal transformations, and conformal symmetry lay at the foundations of Weyl geometry. The conformal symmetry is defined through the requirement of the invariance
with respect to the transformations
\begin{eqnarray}\label{WS}
\text{(i)} \quad \hat g_{\mu\nu}&=&\Sigma^n \,g_{\mu\nu},\qquad \sqrt{-\hat g%
} =\Sigma^{2 n} \sqrt{-g},  \notag \\[5pt]
\text{(ii)} \quad \hat \phi &=& \Sigma^{-n/2} \phi, \qquad
\;\;\hat\psi=\Sigma^{-3n/4}\,\psi,
\end{eqnarray}
where $g=\det g_{\mu\nu}$, $\Sigma(x)>0$, and $\phi$ and $\psi$  describe
real bosonic or fermions fields. In the following we take, without  any loss
of generality, the charge $n=1$.

\paragraph{The Weyl connection.} The natural way to implement the idea of the conformal invariance is to use
the mathematical formalism of Weyl geometry. In a very general approach Weyl
geometry can be defined as the classes of equivalence $(g_{\alpha \beta
},\omega _{\mu }$) of the metric $g_{\mu \nu }$ and the Weyl vector gauge
field ($\omega _{\mu }$), which are related via the Weyl gauge symmetry
transformation (\ref{WS}). These transformations must be supplemented by the
transformation rule of the Weyl vector $\omega _{\mu }$, given by
\begin{equation}\label{WGS}
\hat{\omega}_{\mu }=\omega _{\mu }-\frac{1}{\alpha }\,\partial _{\mu }\ln
\Sigma ,
\end{equation}%
where $\alpha $ denotes the Weyl gauge coupling parameter. A fundamental
property of Weyl geometry is that it is non-metric, with the covariant
derivative of the metric non-vanishing. The non-metricity is determined by
the presence of $\omega _{\mu }$, and it is defined according to $\nu \nu $%
\begin{equation}
\tilde{\nabla}_{\mu }g_{\alpha \beta }=-\alpha \omega _{\mu }g_{\alpha \beta
}.  \label{nm}
\end{equation}%
From Eq.~(\ref{nm}) one obtains the connection $\tilde{\Gamma}$ of the Weyl
geometry as given by
\begin{eqnarray}
\tilde{\Gamma}_{\mu \nu }^{\lambda } &=&\Gamma _{\mu \nu }^{\lambda }+\frac{1%
}{2}\alpha \big[\delta _{\mu }^{\lambda }\,\,\omega _{\nu }+\delta _{\nu
}^{\lambda }\,\,\omega _{\mu }-g_{\mu \nu }\,\omega ^{\lambda }\big]=\Gamma
_{\mu \nu }^{\lambda }+\Psi _{\mu \nu }^{\lambda },  \notag  \label{tGamma}
\\
&&
\end{eqnarray}%
where $\Gamma _{\mu \nu }^{\lambda }$ is the Levi-Civita connection, given
by its usual definition,
\be
\Gamma _{\mu \nu }^{\alpha }(g)=\frac{1}{2}g^{\alpha
\lambda }(\partial _{\mu }g_{\lambda \nu }+\partial _{\nu }g_{\lambda \mu
}-\partial _{\lambda }g_{\mu \nu }),
\ee
and
\be
\Psi _{\mu \nu }^{\lambda
}=\frac{\alpha}{2}\Big(\delta _{\mu }^{\lambda }\,\,\omega _{\nu }+\delta _{\nu
}^{\lambda }\,\,\omega _{\mu }-g_{\mu \nu }\,\omega ^{\lambda }\Big),
\ee
respectively.

Thus, in Weyl geometry, the covariant derivative of a vector
is obtained as
\begin{eqnarray}\label{cov1}
\tilde{\nabla}_{\mu }\omega _{\nu } &=&\frac{\partial \omega _{\nu }}{%
\partial x^{\nu }}-\tilde{\Gamma}_{\mu \nu }^{\lambda }\omega _{\lambda }=%
\frac{\partial \omega _{\nu }}{\partial x^{\nu }}-\Gamma _{\mu \nu
}^{\lambda }\omega _{\lambda }-\Psi _{\mu \nu }^{\lambda }\omega _{\lambda }\nonumber\\
&=&\nabla _{\mu }\omega _{\nu }-\Psi _{\mu \nu }^{\lambda }\omega _{\lambda
}\nonumber\\
&=&\nabla _{\mu }\omega _{\nu }-\alpha \omega _{\mu }\omega _{\nu }+\frac{1}{2%
}\alpha g_{\mu \nu }\omega _{\lambda }\omega ^{\lambda }.
\end{eqnarray}

The contraction of the connection coefficients are given by $\tilde{\Gamma}%
_{\mu \nu }^{\nu }=\tilde{\Gamma}_{\mu }$, and $\Gamma _{\mu \nu }^{\nu
}=\Gamma _{\mu }$, respectively. Then the Weyl vector can be obtained as
\begin{equation*}
\omega _{\mu }=\frac{1}{2\alpha }\left( \tilde{\Gamma}_{\mu }-\Gamma _{\mu
}\right) .
\end{equation*}

Hence, $\omega _{\mu }$ describes the deviation of the trace of the Weyl
connection from the Levi-Civita connection. This also indicates that since $%
\omega _{\mu }$ is part of the Weyl connection $\tilde{\Gamma}$, and thus it has
a purely geometric origin. For the covariant divergence of a vector in the
Weyl geometry we obtain the relation
\begin{equation}\label{cov2}
\tilde{\nabla}_{\lambda }\omega ^{\lambda }=\nabla _{\lambda }\omega
^{\lambda }+2\alpha \omega _{\lambda }\omega ^{\lambda }.
\end{equation}

The Weyl connection $\tilde{\Gamma}$ is invariant with respect to the
combined set of transformations (\ref{WS}) and (\ref{WGS}). If $\omega _{\mu
}$ decouples, $\omega _{\mu }=0$, or it can be represented in a pure gauge form, then the corresponding
geometry is called the Weyl integrable geometry. The Weyl integrable geometry is a metric geometry.

\paragraph{Curvatures in Weyl geometry.} The field strength $F_{\mu\nu}$ of the Weyl vector $\omega_\mu$, is defined
according to
\begin{equation}  \label{W}
\tilde{F}_{\mu\nu} = \tilde{\nabla}_{\mu} \omega_{\nu} - \tilde{\nabla}
_{\nu} \omega_{\mu} = \nabla _{\mu}\omega _\nu-\nabla _\nu \omega _\mu.
\end{equation}

With the help of the Weyl connection $\tilde\Gamma$ one can obtain the
tensor curvatures and the curvature scalar of Weyl geometry, The definitions
are similar to the ones used in the Riemannian geometry, but with $\Gamma $
replaced with $\tilde\Gamma$. Thus, we obtain
\begin{eqnarray}
\tilde R^\lambda_{\mu\nu\sigma}&=& \partial_\nu
\tilde\Gamma^\lambda_{\mu\sigma} -\partial_\sigma
\tilde\Gamma^\lambda_{\mu\nu}
+\tilde\Gamma^\lambda_{\nu\rho}\,\tilde\Gamma_{\mu\sigma}^\rho
-\tilde\Gamma_{\sigma\rho}^\lambda\,\tilde\Gamma_{\mu\nu}^\rho,  \notag \\
\tilde R_{\mu\nu}&=&\tilde R^\lambda_{\mu\lambda\sigma}, \tilde
R=g^{\mu\sigma}\,\tilde R_{\mu\sigma}.
\end{eqnarray}

Explicitly, one finds
\begin{eqnarray}  \label{tRmunu}
\tilde R_{\mu\nu}&=&R_{\mu\nu} +\frac {1}{2}\alpha \left(\nabla_\mu \omega
_\nu-3\,\nabla_\nu \omega _\mu - g_{\mu\nu}\,\nabla_\lambda \omega
^\lambda\right)  \notag \\
&&+\frac{1}{2} \alpha ^2 (\omega _\mu \omega _\nu -g_{\mu\nu}\,\omega
_\lambda \omega ^\lambda),
\end{eqnarray}
\begin{equation}
\hspace{-3.8cm}\tilde R_{\mu\nu}-\tilde R_{\nu\mu}=2 \alpha F_{\mu\nu},
\end{equation}
\begin{equation}\label{tR}
\hspace{-1.5cm}\tilde R= R-3 \,\alpha\,\nabla_\lambda\omega ^\lambda-\frac{3
}{2}\alpha ^2 \omega _\lambda \omega ^\lambda,
\end{equation}
where $R_{\mu \nu}$ and $R$ are the Ricci curvature tensor, and the Ricci
scalar curvature. It is important to note that the right hand sides of the
above equations are in the Riemannian geometry, with the covariant
derivative defined in a standard way as $\nabla_\mu\omega
^\lambda=\partial_\mu \omega ^\lambda+\Gamma^\lambda_{\mu\rho}\,\omega ^\rho$
.
$\tilde R$ has the important property that it transforms covariantly under
the conformal transformations, $\hat{\tilde R}=(1/\Sigma^n)\,\tilde R$.

\paragraph{Conformally invariant gravitational theories.} In Weyl geometry, the simplest gravitational Lagrangian density, satisfying the requirement of  conformal invariance, can be constructed according to the prescription \cite{Weyl,Gh1, Gh2, Gh3,Gh4,Gh4a, Gh4b, Gh5,Gh5a, Gh6, Gh7, Gh8}
\begin{equation}\label{inA}
L_{Weyl}=\left( \frac{1}{4! \, \xi^2} \tilde R^2 - \frac{1}{4} \tilde{F}_{\mu\nu} \tilde{F}^{\mu\nu} \right) \sqrt{-\tilde{g}},
\end{equation}
where  by $\xi < 1$ we have denoted the parameter of the perturbative coupling. The Lagrangian $L_{Weyl}$ can be easily linearized with the use of the substitution  $\tilde{R}^2\rightarrow 2 \phi_0^2\,\tilde R-\phi_0^4$, where $\phi_0$ is an auxiliary scalar field \cite{Gh3,Gh4,Gh5,Gh6}. The new linearized Lagrangian density is mathematically equivalent to the initial one, as one can immediately see from the substitution of the solution $\phi_0^2=\tilde R$ of the equation of motion of $\phi_0$ in the new $L_{Weyl}$. Hence, after performing the linearization procedure, we arrive at a new Lagrangian containing a scalar degree of freedom, given by
\begin{equation}\label{alt3}
L_{Weyl}= \left( \frac{\phi_0^2}{12 \xi^2}  \tilde R - \frac{\phi_0^4}{4!\,\xi^2} - \frac{1}{4} \tilde{F}_{\mu\nu} \tilde{F}^{\mu\nu}  \right) \sqrt{-\tilde{g}}.
\end{equation}

In this Lagrangian density containing the Weyl gauge symmetry, as well as conformal invariance are fully implemented. $L_{Weyl}$ has a spontaneous breaking to an Einstein-Proca Lagrangian of the Weyl gauge field \cite{Gh3,Gh4,Gh5}. Substituting into (\ref{alt3}) the expression of $\tilde{R}$ given in (\ref{tR}), we obtain a Riemann geometric action, invariant under conformal transformation and given by \cite{Gh3,Gh4,Gh5}
\bea\label{a3}
S_{Weyl} &=& \int \Bigg[ \frac{\phi^2}{12 \xi^2} \Big( R - 3\alpha\nabla_\mu \omega^\mu - \frac{3}{2} \alpha^2 \omega_\mu \omega^\mu \Big) \nonumber\\
&&- \frac{ \phi^4}{4! \,\xi^2} - \frac{1}{4} \tilde{F}_{\mu\nu} \tilde{F}^{\mu\nu} \Bigg]  \sqrt{-g} \, d^4x,
\eea
Note that the above Lagrangean has been obtained after performing a gauge transformation, as well as a redefinition of the physical and geometric variables.

\subsection{The Einstein-Hilbert variational principle, and the boundary term}

We consider now the standard gravitational model of Riemann geometry,
constructed with the help of the Ricci scalar, and given by
\begin{equation}\label{act1}
S_{g}=-\frac{1}{2\kappa^2}\int_{\Omega }{R(g)\sqrt{-g}d^{4}x},
\end{equation}
where we have denoted $\kappa^{2}=8\pi G/c^4$. The variation of the Ricci
scalar is given by,
\begin{eqnarray}
\delta \left( R\sqrt{-g}\right) &=&\delta \left( R_{\mu \nu }g^{\mu \nu }
\sqrt{-g}\right)  \notag \\
&=&\left( R_{\mu \nu }-\frac{1}{2}g_{\mu \nu }R\right) \sqrt{-g}\delta
g^{\mu \nu }
+g^{\mu \nu }\delta R_{\mu \nu }\sqrt{-g}. \nonumber\\
\end{eqnarray}

The variation of $R_{\mu \nu }$ can be written as \cite{LaLi} (see Appendix~\ref{appa} for the derivation of this result)
\begin{equation}
\delta R_{\mu \nu }=\nabla _{\lambda }\delta \Gamma _{\mu \nu }^{\lambda
}-\nabla _{\nu }\delta \Gamma _{\mu \sigma }^{\sigma },
\end{equation}
giving
\begin{equation}
g^{\mu \nu }\delta R_{\mu \nu }=g^{\mu \nu }\nabla _{\lambda }\delta \Gamma
_{\mu \nu }^{\lambda }-g^{\mu \lambda }\nabla _{\lambda }\delta \Gamma _{\mu
\sigma }^{\sigma }.
\end{equation}

Hence, we obtain for the variation of the Einstein- Hilbert action the
general expression
\bea\label{gravact}
	\delta S_{g}&=&-\frac{1}{2\kappa^2}\int_\Omega \Big( G_{\mu \nu }\delta g^{\mu \nu }+g^{\mu
		\nu }\nabla _{\lambda }\delta \Gamma _{\mu \nu }^{\lambda }-g^{\mu \lambda
	}\nabla _{\lambda }\delta \Gamma _{\mu \sigma }^{\sigma }\Big)\nonumber\\
 &&\times \sqrt{-g}d^{4}x,
\eea
where we have denoted $G_{\mu \nu }=R_{\mu \nu }-(1/2)Rg_{\mu \nu }$.

As for the variation of the Christoffel symbols, they are given in a
covariant form by the expression
\begin{equation}
\delta \Gamma _{\mu \nu }^{\lambda }=\frac{1}{2}g^{\sigma \lambda }\left[
\nabla _{\nu }\delta g_{\mu \sigma }+\nabla _{\mu }\delta g_{\nu \sigma
}-\nabla _{\sigma }\delta g_{\mu \nu }\right] .  \label{WC}
\end{equation}

For the derivation of the above relation see Appendix~\ref{appb}.

In general relativity, in its standard Riemannian metric formulation, the
boundary term $g^{\mu \nu }\delta R_{\mu \nu }\sqrt{-g}$ is eliminated by
using the Gauss' theorem, extending the surface to infinity, and finally
assuming that the variations of the metric tensor vanish on $\partial \Omega
$,
\begin{equation}
\int_{\Omega }g^{\mu \nu }\delta R_{\mu \nu }\sqrt{-g}d^{4}x=\int_{\Omega
}\nabla _{\lambda }w^{\lambda }\sqrt{-g}d^{4}x=\int_{\partial \Omega
}w^{\lambda }dS_{\lambda },
\end{equation}%
where $dS_{\lambda }$ denotes the element of integration on the hypersurface
$S$ encompassing the four-volume element $d\Omega $.

\section{Gravitational field equations in the presence of Weyl type boundary terms}

We will extend now the gravitational action (\ref{act1}) in the presence of a boundary  to the Weylian
geometric framework, assumed to be realized on the boundary $\Omega$ of the space-time.  Thus, we abandon the assumption of the metricity of the
geometry on $\Omega$, and thus near and on the boundary we replace the covariant derivative of the metric tensor
by its expression (\ref{nm}), so that $\nabla \rightarrow \tilde{\nabla}$,
while the Christoffel connection, and its variation, are replaced by the
Weyl connection, $\Gamma \rightarrow \tilde{\Gamma}$, and by $\delta \Gamma
\rightarrow \delta \tilde{\Gamma}$. Therefore, in calculating the variation of the action, and in the final field equation,  the Riemannian boundary is replaced by its Weylian counterpart.

We will consider now the variation of the action (\ref{act1}) by assuming the presence of the boundary terms in the integral. By varying the action with respect to the metric we obtain
\bea\label{25}
\delta S_g&=&-\frac{1}{2\kappa ^2}\int_\Omega\left(R_{\mu \nu}-\frac{1}{2}Rg_{\mu \nu}\right)\delta g^{\mu\nu}\sqrt{-g}d^4x\nonumber\\
&&-\frac{1}{2\kappa ^2}\int_\Omega {\left(g^{\mu
		\nu }\nabla _{\lambda }\delta \Gamma _{\mu \nu }^{\lambda }-g^{\mu \lambda
	}\nabla _{\lambda }\delta \Gamma _{\mu \sigma }^{\sigma }\right)\sqrt{-g}d^4x}.\nonumber\\
\eea

In the following we will investigate the action (\ref{25}) for a particular choice of the boundary term, in which the Christoffel symbols, as well as their variations, are redefined in a Weyl geometric framework. We would like to point out that due to the arbitrariness in the choice of the boundary terms, such an approach is always possible. Essentially, our approach consists in substituting in the standard expressions of the variations of the Christoffel symbols the Riemannian covariant derivative with the Weylian one.

\subsection{Variation of the boundary terms}

We note first that from the condition $g_{\mu \nu }g^{\nu \sigma }=\delta
_{\mu }^{\sigma }$, we obtain
\begin{equation}
\tilde{\nabla}_{\lambda }g^{\mu \nu }=\alpha \omega _{\lambda }g^{\mu \nu },
\end{equation}
and
\begin{equation}
\delta g_{\mu \lambda }=-g_{\mu \nu }g_{\lambda \sigma }\delta g^{\nu \sigma
},
\end{equation}
respectively. Moreover, by taking into account the smallness of the Weyl
vector, and of the variation of the metric tensor, \textit{we assume} that
\begin{equation}
\tilde{\nabla}_{\sigma }\delta g_{\mu \nu }=-\alpha \omega _{\sigma }\delta
g_{\mu \nu }=\alpha \omega _{\sigma }g_{\gamma \mu }g_{\rho \nu }\delta
g^{\gamma \rho }.
\end{equation}
\newline

For the variation of the boundary Weyl connection, defined according to

\begin{equation}
\delta \tilde{\Gamma}_{\mu \nu }^{\lambda }=\frac{1}{2}g^{\sigma \lambda }%
\left[ \tilde{\nabla}_{\nu }\delta g_{\mu \sigma }+\tilde{\nabla}_{\mu
}\delta g_{\nu \sigma }-\tilde{\nabla}_{\sigma }\delta g_{\mu \nu }\right] ,
\label{WC1}
\end{equation}%
we find, with the use of Eq. (\ref{nm}), the expression
\begin{equation}
\delta \tilde{\Gamma}_{\mu \nu }^{\lambda }=\frac{\alpha }{2}\left( \omega
_{\nu }g_{\mu \rho }\delta _{\gamma }^{\lambda }+\omega _{\mu }g_{\nu \gamma
}\delta _{\rho }^{\lambda }-\omega ^{\lambda }g_{\mu \gamma }g_{\nu \rho
}\right) \delta g^{\rho \gamma }.
\end{equation}

Hence, the first boundary term can be obtained as
\begin{eqnarray}
\hspace{-0.9cm}\tilde{\nabla}_{\lambda }\delta \tilde{\Gamma}_{\mu \nu
}^{\lambda } &=&\frac{\alpha }{2}\Bigg[\left( \tilde{\nabla}_{\gamma }\omega
_{\nu }\right) g_{\mu \rho }+\left( \tilde{\nabla}_{\rho }\omega _{\mu
}\right) g_{\nu \gamma }  \notag \\
\hspace{-0.9cm} &&-\left( \tilde{\nabla}_{\lambda }\omega ^{\lambda }\right)
g_{\mu \gamma }g_{\nu \rho }+\alpha g_{\mu \gamma }g_{\nu \rho }\omega
^{\lambda }\omega _{\lambda }\Bigg]\delta g^{\rho \gamma },
\end{eqnarray}%
and
\begin{eqnarray}
g^{\mu \nu }\tilde{\nabla}_{\lambda }\delta \tilde{\Gamma}_{\mu \nu
}^{\lambda }&=&\frac{\alpha }{2}\Big[\tilde{\nabla}_{\gamma }\omega _{\rho }
+\tilde{\nabla}_{\rho }\omega _{\gamma } - g_{\gamma \rho}\tilde{\nabla}%
_{\lambda }\omega ^{\lambda }  \notag \\
&&+\alpha g_{\gamma \rho }\omega ^{\lambda }\omega _{\lambda }\Big] \delta
g^{\rho \gamma },
\end{eqnarray}
respectively. For the second boundary term we find successively
\begin{equation}
\delta \tilde{\Gamma}_{\mu \sigma }^{\sigma }=\frac{\alpha }{2}\left( \omega
_{\gamma }g_{\mu \rho }+\omega _{\mu }g_{\gamma \rho }-\omega _{\rho }g_{\mu
\gamma }\right) \delta g^{\rho \gamma },
\end{equation}
\begin{equation}
\tilde{\nabla}_{\lambda }\delta \tilde{\Gamma}_{\mu \sigma }^{\sigma }=\frac{
\alpha }{2}[(\tilde{\nabla}_{\lambda }\omega _{\rho })g_{\mu \gamma }+(%
\tilde{\nabla}_{\lambda }\omega _{\mu })g_{\gamma \rho }-(\tilde{\nabla}%
_{\lambda }\omega _{\rho })g_{\mu \gamma }]\delta g^{\rho \gamma },
\end{equation}
and
\begin{eqnarray}
g^{\mu \lambda }\tilde{\nabla}_{\lambda }\delta \tilde{\Gamma}_{\mu \sigma
}^{\sigma } &=&\frac{\alpha }{2}\left[ \Big(\tilde{\nabla}_{\rho }\omega
_{\gamma }\right) +g^{\mu \lambda }\left( \tilde{\nabla}_{\lambda }\omega
_{\mu }\right) g_{\gamma \rho }  \notag \\
&&-\left( \tilde{\nabla}_{\gamma }\omega _{\rho }\right) \Big]\delta g^{\rho
\gamma },
\end{eqnarray}%
respectively. Therefore for the total variation of the Weylian boundary
terms we finally obtain
\begin{eqnarray}\label{40}
&&g^{\mu \nu }\tilde{\nabla}_{\lambda }\delta \tilde{\Gamma}%
_{\mu \nu }^{\lambda }-g^{\mu \lambda }\tilde{\nabla}_{\lambda }\delta
\tilde{\Gamma}_{\mu \sigma }^{\sigma }=\frac{\alpha }{2}\times \nonumber\\
&& \Big[\Big(2\tilde{%
\nabla}_{\gamma }\omega _{\rho }-g_{\gamma \rho }\tilde{\nabla}_{\lambda
}\omega ^{\lambda }
-g^{\mu \lambda }g_{\gamma \rho }\tilde{\nabla}_{\lambda
}\omega _{\mu }\Big)
+\alpha \omega ^{\lambda }\omega
_{\lambda }g_{\gamma \rho }\Big]\delta g^{\rho \gamma }.\nonumber\\
\end{eqnarray}

\subsection{Action and field equations in the presence of Weyl type boundary terms}

Thus, with the use of Eq.~(\ref{40}), we  arrive at the result that the variation of the Einstein-Hilbert gravitational action
 in the presence of Weyl type boundary terms takes the form
\begin{eqnarray}
\hspace{-0.6cm}&&\delta \tilde{S}_{g} = -\frac{1}{2\kappa^2} \int_{\Omega }\Bigg[ G_{\mu \nu }+\frac{\alpha ^2}{2}g_{\mu \nu}\omega_\lambda \omega ^\lambda+\frac{%
\alpha }{2} \times \nonumber\\
\hspace{-0.6cm}&&\left( 2\tilde{\nabla}_{\mu }\omega _{\nu }-g_{\mu \nu }\tilde{%
\nabla}_{\lambda }\omega ^{\lambda }-g^{ \sigma \lambda }g_{\mu \nu }\tilde{%
\nabla}_{\lambda }\omega _{\sigma }\right)
\Bigg] \delta g^{\mu \nu }\sqrt{-g}%
d^{4}x.
\end{eqnarray}%

With the help of the expressions of the Weyl covariant derivatives as given by Eqs.~(\ref{cov1}) and (\ref{cov2}), we obtain the gravitational field equations in the presence of Weylian boundary terms, and in the absence of matter,  as given by
\begin{eqnarray}\label{feqa}
G_{\mu \nu }+\alpha \left(\nabla _\mu \omega _\nu-g_{\mu \nu}\nabla _\lambda \omega ^\lambda\right)-\alpha ^2\left(\omega _\mu \omega _\nu+\frac{1}{2}\omega ^2 g_{\mu \nu}\right)=0, \nonumber\\
\end{eqnarray}
where we have denoted $\omega ^2=\omega _\lambda \omega ^\lambda$.

 The field equations (\ref{feqa}) are defined in a metric Riemann geometry. The presence of the boundary terms is equivalent to the introduction in  the field equations of a new term, defined in a Weyl geometric framework. In the following Section we will consider an application of this mathematical formalism, by discussing the effects of the Weyl type boundary effects on the evolution of the Universe in  the warm inflationary scenario.

\section{Warm inflation in the presence of Weylian boundary terms} \label{action}

In the present Section we will introduce a warm inflationary model in which the gravitational field equations
also contain the non-vanishing boundary terms discussed in the previous Section, and a dissipative
cosmological scalar field. In order to obtain a consistent description of the cosmological evolution,  we
will also formulate the dissipative dynamics of the scalar field in terms of
an action principle, by adding the Lagrangian of the dissipative scalar to the total Lagrangian. The boundary terms are included in the gravitational
field equations by considering the geometric extension from a Riemann to a
Weyl type geometry. We also assume that the gravitational action, linear in the Ricci scalar, is the result of the symmetry breaking of the linearized Weyl geometric action, with the scalar field substituted by its vacuum expectation value. Thus, the scalar field is decoupled from geometry, and the presence of the scalar fields in the early Universe, having a geometric nature, is assumed to be a consequence of the symmetry breaking of the conformal invariance of a higher order action.

\subsection{Standard warm inflation}

In the warm inflation scenario, the dynamical evolution of the very early Universe is dominated by the interaction between the
inflaton field, and other physical fields, via dissipative effects. The energy
dissipation occurs through various physical processes, such as particle
scattering, decays, and thermal radiation, which transfer energy from the
inflaton field to other fields, and ultimately, to radiation.

To describe the early evolution of the Universe one adopts the standard, flat, homogeneous, and isotropic  Friedmann-Lemaître-Robertson-Walker (FLRW) metric, which is given by
\begin{equation}\label{metr}
	ds^{2}=c^{2}dt^{2}-a^{2}(t)\left( dx^{2}+dy^{2}+dz^{2}\right) .
\end{equation}
 where $a(t)$ represents the scale factor of the Universe. Moreover, we also introduce a fundamental cosmological quantity, the Hubble function, given by $H=\dot{a}/a$, where a dot denotes the derivative with respect to the cosmological time $t$.

The dynamical evolution of the Universe during the warm inflationary epoch is described
by the Friedmann equations \cite{BF,B}, given by
\begin{eqnarray}
3H^{2} &=& c^4 \kappa^2 \rho _{rad}+\rho _{\phi },\\
2\dot{H} + 3H^2 &=& -c^2 \kappa^2 p_{rad}+ p_{\phi},
\end{eqnarray}
where $\kappa^2=8\pi G/c^4$, $G$ is the gravitational constant,  $\rho _{\phi }$, $\rho _{rad}$ are the energy densities of the scalar
field and radiation, respectively, while $p_{\phi }$, $p_{rad}$ are the pressures of the radiation fluid, and of the scalar field.  The
Friedmann equations relate the expansion rate of the Universe, given by the
Hubble function $H$, to the energy density and pressure of the scalar field
and radiation, respectively.

The thermal effects due to interactions between the scalar field and other
particles are described by the energy balance equations, given by
\begin{eqnarray}
\dot{\rho}_{\phi }+3H\left( \rho _{\phi }+p_{\phi }\right) &=&-\Gamma \dot{%
\phi}^{2} \\
\dot{\rho}_{rad}+3H\left( \rho _{rad}+p_{rad}\right) &=&\Gamma \dot{\phi}%
^{2},
\end{eqnarray}%
where $\Gamma $ is the dissipation coefficient.

The evolution of the scalar field is given by a Klein-Gordon type equation,
\begin{equation}
\ddot{\phi}+3H(1+Q)\dot{\phi}+V^{\prime }(\phi )=0,
\end{equation}%
where the coefficient $Q=\Gamma /3H$ designates the ratio of thermal damping
to expansion damping \cite{H}. By using the above equations, one can study the generation of the radiation in the early Universe, together with the corresponding cosmological dynamics. Hence, in warm inflation, there is no need for a post-inflationary phase (reheating), during which matter is generated from a Universe cooled down by the de Sitter type expansion. The standard Big Bang scenario is thus recovered, and the isotropy and homogeneity problems are solved simultaneously during a matter creation phase.

\subsection{The gravitational field equations in the presence of a dissipative scalar field and Weyl geometric boundary terms}

We consider an action that describes the early Universe which includes three distinct terms,
and is given by
\begin{eqnarray}\label{S}
S &=& -\frac{1}{2\kappa^2}\int _\Omega{R(g)\sqrt{-g}d^{4}x} + \frac{1}{\kappa ^2} \int{\mathcal{L}_{\phi}\sqrt{-g}%
d^{4}x} \nonumber\\
&&+ \int{\mathcal{L}_{m}\sqrt{-g}d^{4}x}.
\end{eqnarray}

The first term arises from the standard description of the gravitational field, and its variation with respect to the metric was calculated
in the previous Section, with the Weyl type boundary effects also included. This term captures the influence of the spacetime
geometry, and curvature, on the dynamics of the early Universe. The second term accounts
for dissipative effects related to the presence of a scalar field, and represents the processes that, on one hand, trigger the accelerated expansion of the Universe, and, on the other hand, lead to the loss of
the energy of the scalar field, and to the creation of matter and radiation in the evolving Universe. We assume that the scalar field is of geometric origin, and it has its source in the integrable Weyl geometry, originating from the Weyl vector $\omega _\mu$. Finally, the third term represents the
contribution of the newly created matter to the action.

\subsubsection{The generalized dissipative Klein-Gordon equation}

To describe the radiation generation in the early Universe we adopt for the Lagrange density of the dissipative scalar field  the following form \cite{dis}
\begin{equation}
\mathcal{L}_{\phi }= e^{\Gamma(\phi)} \left[ \frac{1}{2}g^{\mu \nu } \frac{%
\partial \phi }{ \partial x^{\mu }}\frac{\partial \phi }{\partial x^{\nu }}%
-V\left( \phi\right) \right]\sqrt{-g},  \label{L}
\end{equation}
where $-g$ is the determinant of the covariant background metric tensor $%
g_{\mu \nu}$. In the next Sections we will assume for $\Gamma$ the simple form $\Gamma = \beta \phi$, where $\beta $ is a constant.

The Euler-Lagrange equations, giving the minimum of the action of the dissipative scalar field, are
\begin{equation}
	\partial _{\mu }\frac{\partial \mathcal{L}_{\phi }}{\partial \phi _{,\mu }}-%
	\frac{\partial \mathcal{L}_{\phi }}{\partial \phi }=0. \label{lagrange}
\end{equation}

Therefore, given our dissipative  Lagrangian, we obtain
\begin{equation}
	\frac{\partial }{\partial x^{\mu }}\left( \sqrt{-g} e^{\beta \phi} g^{\mu \nu }\frac{%
		\partial \phi }{\partial x^{\nu }}\right) +\frac{\partial}{\partial \phi}\left(\sqrt{-g} e^{\beta \phi}V(\phi)\right)=0.
\end{equation}

It is useful to note that
\begin{equation}
	\frac{1}{\sqrt{-g}}\frac{\partial }{\partial x^{\mu }}\left( \sqrt{-g}%
	g^{\mu \nu }\frac{\partial \phi }{\partial x^{\nu }}\right) =\nabla
	_{\mu }\nabla ^{\mu }\phi =\Box \phi, \label{box}
\end{equation}
and therefore, we can rewrite Eq. (\ref{lagrange}) in the following form
\begin{equation}\label{eom}
	\Box\phi + \frac{dV(\phi)}{d\phi} + \beta \left[g^{\mu \nu}\partial_{\mu}\phi \partial_{\nu}\phi+V(\phi)\right] = 0.
\end{equation}

Eq.~(\ref{eom}) gives the evolution of a scalar field in the presence of dissipation, with the dissipation function $\Gamma$ having a simple form, proportional to the scalar field.

\subsubsection{The gravitational field equations}

By varying the action (\ref{S}) with respect to the metric tensor we get the
equations of motion for the metric tensor, the scalar field, and matter in a Universe with
boundary. The system of equations, generalizing the standard Einstein equations, are thus obtained as
\bea
&&R_{\mu \nu}-\frac{1}{2}Rg_{\mu \nu}+\alpha \left(\nabla _\mu \omega _\nu-g_{\mu \nu}\nabla _\lambda \omega ^\lambda\right)\nonumber\\
&&-\alpha ^2\left(\omega _\mu \omega _\nu+\frac{1}{2}\omega ^2 g_{\mu \nu}\right)= \kappa^2 T_{\mu \nu}^{(m)}+ T^{(\phi)}_{\mu \nu},
\eea
where
\be
T_{\mu \nu}^{(m)} = \frac{2}{\sqrt{-g}}\frac{\partial (\sqrt{-g}\mathcal{L}%
_{m})}{\partial g^{\mu \nu }},
\ee
is the well known energy-momentum tensor of the matter component, and $T^{(\phi)}_{\mu \nu}$ is the energy-momentum tensor of the scalar field. We assume now that the Weyl vector can be expressed as the gradient of a scalar function $\psi$, so that $\omega _\mu =\nabla _\mu \psi$. Hence, with this assumption, it follows that the space-time boundary could be described by the integrable Weyl geometry. Moreover, we suppose that the Weyl geometric effects are small, and hence the terms $\omega ^2/2$, and $\nabla _\mu \nabla _\nu \psi$ can be safely ignored in the field equations. By taking into account that $\nabla _\lambda \omega ^\lambda=\nabla _\lambda \nabla ^\lambda \psi=\Box \psi$, we can reformulate the field equations as
	\begin{eqnarray}
		 R_{\mu \nu}-\frac{1}{2}Rg_{\mu \nu} - \alpha \Bigg( \partial_{\mu}\psi \partial_{\nu}\psi +\alpha g_{\mu \nu}\Box \psi \Bigg)= \kappa^2 T_{\mu \nu}^{(m)}+ T^{(\phi)}_{\mu \nu}.\nonumber\\
	\end{eqnarray}
To simplify the mathematical formalism we rescale the Weyl scalar as $\psi \rightarrow \alpha \psi$, and $\alpha ^3 \rightarrow \alpha ^2/2$. Then, we can write the gravitational field equations in the form
	\begin{eqnarray}\label{fe}
		 R_{\mu \nu}-\frac{1}{2}Rg_{\mu \nu} - \frac{\alpha ^2}{2} \Bigg( \partial_{\mu}\psi \partial_{\nu}\psi +g_{\mu \nu}\Box \psi \Bigg)= \kappa^2 T_{\mu \nu}^{(m)}+ T^{(\phi)}_{\mu \nu}.\nonumber\\
	\end{eqnarray}

The energy-momentum tensor of the dissipative scalar field has the following form
\begin{eqnarray}
\hspace{-0.5cm}T_{\mu \nu }^{(\phi)}&=&\frac{2}{\sqrt{-g}}\frac{\delta S_{\phi }}{\delta
g^{\mu \nu }} = 2\frac{\partial \mathcal{L}_{\phi}}{\partial g^{\mu
\nu}} - g_{\mu \nu}\mathcal{L}_{\phi}  \notag \\
\hspace{-0.5cm}&=& e^{\Gamma} \left[\partial_{\mu}\phi \partial_{\nu}\phi-g_{\mu \nu }\left( \frac{1}{2}g^{\sigma \rho }\partial_{\sigma}\phi
\partial_{\rho}\phi-V(\phi)\right)\right].
\end{eqnarray}

The energy-momentum tensor for the dissipative scalar field can be represented in a form similar to the standard matter form,
after introducing the energy density and the pressure of the scalar field
\begin{equation}
	T_{\alpha \beta} ^{(\phi)}= (p_{\phi}+\rho_{\phi})u_{\alpha}u_{\beta}- p_{\phi}g_{\alpha \beta},
\end{equation}
where
\begin{eqnarray}\label{p_phi}
	\rho_{\phi} &=& e^{\Gamma}\left( \frac{1}{2}g^{\mu \nu }\phi _{,\mu }\phi
	_{,\nu }+V(\phi) \right), \\
p_{\phi} &=& e^{\Gamma}\left( \frac{1}{2}g^{\mu \nu }\phi _{,\mu }\phi _{,\nu
	}-V(\phi) \right), \\
	u_{\mu} &=& \frac{\phi_{,\mu}}{\sqrt{g^{\alpha \beta }\phi _{,\alpha }\phi_{,\beta}}},
\end{eqnarray}
respectively.

The energy-momentum tensor for the matter is obtained similarly through the pressure and density of matter
\begin{equation}
	T_{\mu \nu}^{(m)}=\left(\rho_m+p_m\right)u_\mu u_\nu -p_m g_{\mu \nu}.
\end{equation}

\subsection{Friedmann equations in a Weyl geometric  Universe with boundary}

We will apply now the general equations derived in the previous Sections to the specific case of the FLRW metric, assumed to also describe the very early Universe.

The Lagrangian describing the dissipative cosmological evolution of a scalar field
in the early Universe with metric (\ref{metr}) is given by
\begin{equation}
	L = e^{\beta\phi}\left[\frac{1}{2}\dot{\phi}^2 -V(\phi)\right]a^3,
\end{equation}
where $\sqrt{-g} = a^3(t)$. The dot stands for the first derivative with respect to time, $%
\dot{\phi} = d \phi/dt$, as the scalar field is a function of time only, $%
\phi=\phi(t)$.

Thus, in the FLRW geometry, the generalized Klein-Gordon equation  Eq.~(\ref{eom}) becomes
\begin{equation}\label{51}
	\ddot{\phi} + 3H\dot{\phi} + V^{\prime}(\phi) = -\beta \dot{\phi}^2 - \beta V(\phi),
\end{equation}
with the notation $V^{\prime}$ denoting the first derivative of the potential with respect to the scalar field. The Klein-Gordon equation can be reformulated as
\be\label{Q1}
\ddot{\phi}+3H(1+Q)\dot{\phi}+V'(\phi)=0,
\ee
where
\be\label{Q2}
Q=\frac{\beta}{3H}\left[ \dot{\phi}+\frac{ V(\phi)}{\dot{\phi}}\right]=\frac{\Gamma}{3H}.
\ee

The first Friedmann equation is derived from the field equations (\ref{fe}), and it is given by
\begin{eqnarray}\label{52}
	3H^2 &=& 8\pi G \rho_m+e^{\beta \phi}\left(\frac{\dot{\phi}^2}{2}+V(\phi)\right)\nonumber\\
&&+\frac{\alpha^2}{2}\left(\ddot{\psi}+3H\dot{\psi}+\dot{\psi}^2\right)=\rho_{eff}.
\end{eqnarray}

We obtain the second Friedmann equation in a similar manner as
\begin{eqnarray}\label{53}
	2\dot{H} + 3H^2 &=& -\frac{8\pi G}{c^2}p_m- e^{\beta \phi}\left(\frac{\dot{\phi}^2}{2}-V(\phi)\right)\nonumber\\
&&+\frac{\alpha^2}{2}\left(\ddot{\psi}+3H\dot{\psi}\right)=-p_{eff}.
\end{eqnarray}

Eqs.~(\ref{51}), (\ref{52}) and (\ref{53}) represent the basic equations of our cosmological model. If we substitute the first Friedmann equation into the second one, we obtain the following relation for the derivative of the Hubble function
\begin{equation}
\dot{H}=-4\pi G\left(\rho_m+\frac{p_m}{c^2}\right)-e^{\beta \phi}\frac{\dot{\phi}^2}{2}-\frac{\alpha ^2}{4}\dot{\psi}^2.
\end{equation}

We also immediately obtain the relation
\begin{equation}
\rho_{eff}+p_{eff}=8\pi G \left(\rho_m+\frac{p_m}{c^2}\right)+e^{\beta \phi}\dot{\phi}^2+\frac{\alpha ^2}{2}\dot{\psi}^2.
\end{equation}

\section{Warm inflation with Weylian boundary terms} \label{models}

In the present Section we will investigate specific models of the warm inflationary scenario in Weyl geometric gravity, in the presence of boundary terms.

From Eqs.~(\ref{52}) and (\ref{53}) we obtain the conservation equation
\begin{equation}
\dot{\rho}_{eff}+3H\left(\rho_{eff}+p_{eff}\right)=0.
\end{equation}

Explicitly, the conservation equation takes the form
\begin{eqnarray}\label{72}
&&8\pi G\left[\dot{\rho}_m+3H\left(\rho_m+\frac{p_m}{c^2}\right)\right]-\beta \frac{\dot{\phi}^3}{2}e^{\beta \phi}\nonumber\\
&&+\frac{\alpha ^2}{2}\frac{d}{dt}\left(\ddot{\psi}+3H\dot{\psi}+\dot{\psi}^2\right)+\frac{3\alpha ^2}{2}H\dot{\psi}^2=0.
\end{eqnarray}

Eq.~(\ref{72}) is the basic equation of the present warm inflationary model. It describes the creation of matter from two scalar fields, $\phi$, has its origin in the linearization and the symmetry breaking of the initial conformally invariant Weyl action. The second scalar field $\psi$ describes the contribution of the boundary terms, interpreted in a Weyl geometric framework. Furthermore, we assume that these two scalar fields, having different geometrical physical origins, are distinct, even that the possibility that $\phi =\psi$ could also lead to viable cosmological models.

The global energy balance equation Eq.~(\ref{72}) allows the possibility of the existence of several scenarios for radiation creation from the two scalar fields, depending on the splitting of the conservation equation. Once the matter creation/scalar field dissipation model is adopted, the system of cosmological equations describing evolution of the early Universe is closed, and it can be analyzed by using analytical or numerical techniques. In the following we will investigate three such scenarios, corresponding to three distinct models describing radiation creation in the early Universe.

\subsection{Model I: $\ddot{\psi}+3H\dot{\psi}+\dot{\psi}^2=\Lambda$}

As a first warm inflationary model, we consider the splitting of the global conservation equation in the following components
\begin{equation}
8\pi G\left[\dot{\rho}_m+3H\left(\rho_m+\frac{p_m}{c^2}\right)\right]=\beta \frac{\dot{\phi}^3}{2}e^{\beta \phi}-\frac{3\alpha ^2}{2}H\dot{\psi}^2,
\end{equation}
and
\begin{equation}
\ddot{\psi}+3H\dot{\psi}+\dot{\psi}^2=\Lambda={\rm constant}. \label{cons}
\end{equation}

In this model the creation of matter is driven by the scalar field, and its derivative, which gives a positive contribution to the creation rate if the term $\beta \dot{\phi}^3>0$, while the boundary term contributes to a weakening of the creation rate. Particle creation is possible if and only if the condition $\beta \frac{\dot{\phi}^3}{2}e^{\beta \phi}>\frac{3\alpha ^2}{2}H\dot{\psi}^2$ is satisfied during the entire phase of cosmological evolution.

As  for the matter content of the early Universe, we assume that it consisted mainly of radiation, satisfying the equation of state
\begin{equation}
p_m=\frac{\rho_mc^2}{3},
\end{equation}
with
\begin{equation}
\rho_m=\frac{\pi ^2k_B^4}{15c^5\hbar ^3}T^4=\sigma T^4, \label{SB}
\end{equation}
where $k_B$ and $\hbar$ are the Boltzmann and the Planck constants, respectively.

\paragraph{Dimensionless form of the evolution equations.}
To simplify the mathematical formalism we introduce the set of dimensionless variables $(h,\tau,r_m,v(\phi),\lambda)$, defined as
\begin{equation}
H=H_0h,\tau =H_0t,\rho _m=\frac{3H_0^2}{8\pi G}r_m, v(\phi)=\frac{V(\phi)}{H_0^2},\lambda =\frac{\Lambda}{ H_0^2}.
\end{equation}

Then, the basic equations describing the warm inflationary model with the Weyl type boundary conditions satisfying the condition (\ref{cons}) are
\begin{equation}\label{M1}
\frac{d^2\phi}{d\tau ^2}+3h\frac{d\phi}{d\tau}+v'(\phi)=-\beta \left(\frac{d\phi}{d\tau}\right)^2-\beta v(\phi),
\end{equation}
\begin{equation}\label{M2}
\frac{d^2\psi}{d\tau ^2}+3h\frac{d\psi}{d\tau}+\left(\frac{d\psi}{d\tau}\right)^2=\lambda,
\end{equation}
\begin{eqnarray}\label{M3}
2\frac{dh}{d\tau}+3h^2&=&-r_m-e^{\beta \phi}\left[\frac{1}{2}\left(\frac{d\phi}{d\tau}\right)^2-v(\phi)\right]\nonumber\\
&&+\frac{\alpha ^2}{2}\left[\lambda -\left(\frac{d\psi}{d\tau}\right)^2\right],
\end{eqnarray}
\begin{equation}\label{M4}
\frac{dr_m}{d\tau}+4hr_m=\frac{\beta }{6}\left(\frac{d\phi}{d\tau}\right)^3 e^{\beta \phi}-\frac{\alpha ^2}{2}h\left(\frac{d\psi}{d\tau}\right)^2,
\end{equation}

In the following we will investigate the solutions of the above system by considering several functional forms of the scalar field potential. In particular, we will study radiation creation in the presence of a null potential, $v(\phi)=0$,  a quadratic potential of the form $v(\phi)=m\phi ^2/2$, the Higgs type potential $v(\phi)=\mu \phi^2/2+\nu \phi^4/4$, and an exponential one, given by $v(\phi)=v_0e^{-\delta \phi}$.

In addition, certain initial conditions can be determined from the dimensionless form of the first Friedmann equation,
\begin{equation}
	h^2=r_m+\frac{e^{\beta \phi}}{3}\left[\frac{1}{2}\left(\frac{d\phi}{d\tau}\right)^2+v(\phi)\right]+\frac{\alpha^2}{6}\lambda.
\end{equation}
Therefore, the parameters $\alpha, \beta, \lambda$ are related to $\phi '(0)=\phi_{01}$ and $\psi (0)=\psi_0$ by the relation
\begin{equation}
	1-\frac{e^{\beta \phi(0)}}{3}\left[\frac{\phi'(0)^2}{2}+v(\phi(0))\right]=\frac{\alpha^2 \lambda}{6}. \label{c1}
\end{equation}

\paragraph{Numerical results.}
We chose for the initial conditions the following numerical values: $h(0) = 1,\; r_{m}(0) = 0,\; \phi(0) = 4,\; \phi'(0) = -2.5,\; \psi(0) = 2,\; \psi'(0) = 1.5$. For the parameters, the numerical values were selected to satisfy the condition (\ref{c1}), $\alpha = 0.001,\; \beta_1 = -0.0102,\; \beta_2 = -0.0403, \beta_3 = -0.0651,\; \beta_4 = -0.0245,$ and $\lambda = 0.1,$ respectively. The various values for $\beta$ were chosen in accordance to the multiple forms of the scalar potential considered, which are $v_1=0,\; v_2=0.025\phi^2,\; v_3=0.016\phi^2 + 0.002\phi^4$ and $ v_4=0.18e^{-0.01\phi}$, respectively.

\begin{figure*}[htbp]
\includegraphics[scale=0.9]{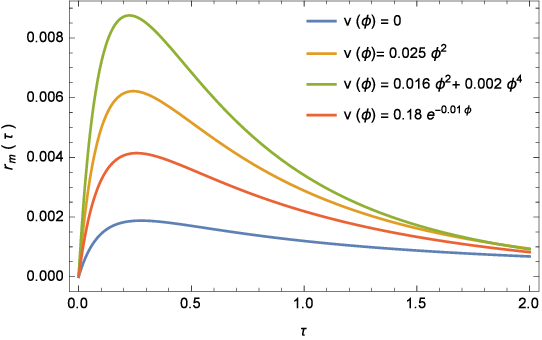}
\includegraphics[scale=0.9]{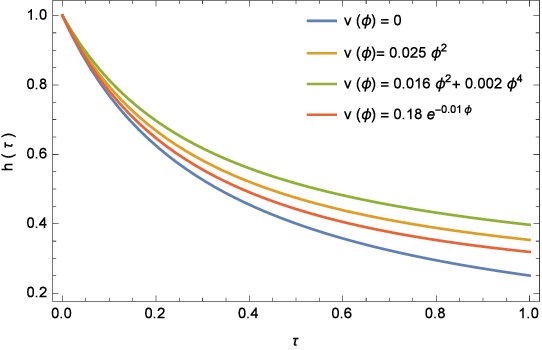}
\caption{Variations as functions of the dimensionless cosmological time of the radiation density $r_m$ (left panel) and of the dimensionless Hubble function $h$ (right panel) for the four considered forms of the scalar field potential in the warm inflationary model with $\ddot{\psi}+3H\dot{\psi}+\dot{\psi}^2=\Lambda$.}\label{fig1}
\end{figure*}

 The time variations of the matter density  $r_m$ and of the Hubble function $h$ are represented in Fig.~(\ref{fig1}). The radiation energy density monotonically increases from the initial zero value until it reaches a maximum value. The time at which the  maximum value of $r_m$ reached is relatively independent of the functional form of the potential, and the maximum value  occurs at $\tau \approx 0.243$. The maximum values of the energy density of the radiation component, corresponding to the various forms of the dimensionless scalar field potential $v(\phi)$, are given in Table~\ref{table1}. The different scalar potentials produce distinct maximum values for the radiation density,

 \begin{table}[h!]
 \begin{center}
\begin{tabular}{|c|c|c|}
\hline
\hline
$v(\phi )$ & $\tau _{\max }$ & $r_{m}^{(\max )}\times 10^{3}$ \\
\hline
$0$ & $0.259$ & $1.85$ \\
\hline
$0.025\phi ^{2}$ & $0.235$ & $6.18$ \\
\hline
$0.016\phi ^{2}+0.002\phi ^{4}$ & $0.229$ & $8.72$ \\
\hline
$0.18e^{-0.01\phi }$ & $0.247$ & $4.11$ \\
\hline
\hline
\end{tabular}
\caption{Maximum values of the radiation energy density for different scalar field potentials for $\ddot{\psi}+3H\dot{\psi}+\dot{\psi}^2=\Lambda$.}\label{table1}
\end{center}
\end{table}

  During this first phase, particle creation processes dominate the overall dynamics, and thus the radiation density increases significantly. Once the maximum of the matter density is reached, the expansion of the Universe becomes dominant, and the particle number density decreases. The Hubble function is a monotonically decreasing function of time, indicating an expansionary evolution of the warm inflationary Universe. The nature of the cosmological evolution is strongly dependent on the functional form of the scalar field potential.

\begin{figure*}[htbp]
\includegraphics[scale=0.9]{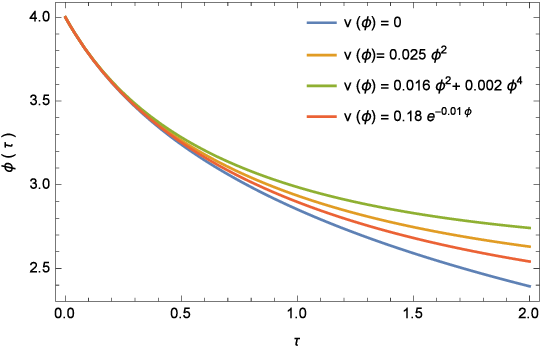}
\includegraphics[scale=0.9]{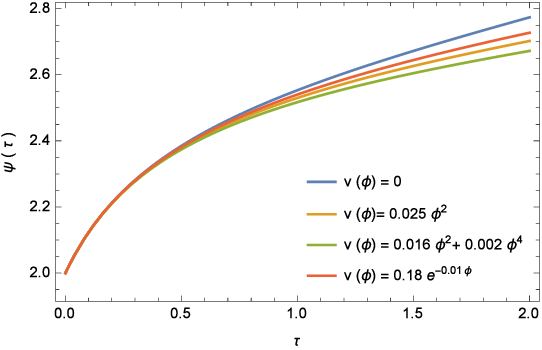}
\caption{Variations as functions of the dimensionless cosmological time of the scalar field  $\phi $ (left panel) and of the boundary contribution  $\psi$ (right panel) for the four considered forms of the scalar field potential in the warm inflationary model with $\ddot{\psi}+3H\dot{\psi}+\dot{\psi}^2=\Lambda$.}\label{fig2}
\end{figure*}

The variations of the scalar field $\phi$, and of the field $\psi$,  are represented in Fig.~(\ref{fig2}). The field $\phi$ decreases rapidly during the inflationary phase. In the time interval $\tau \leq 0.55$, the dynamics of the field is basically independent of the nature of the scalar field potential, but at larger times the evolution of $\phi$ is dependent of the form of the potential. On the other hand, the boundary field $\psi$ increases in time, which suggests that the Weyl boundary effects will persist even after the transition from the inflationary era to the decelerating phase in the evolution of our Universe. Significantly, the behaviour of $\psi$ strongly depends on the parameter $\lambda$.

The behaviour of the energy density of radiation, and of the scalar fields are significantly influenced by the parameters $\alpha$ (derived from the non-metricity condition), $\beta$ (the dissipative constant), and $\lambda$ (accounting for the equation of conservation for the boundary terms in Eq. (\ref{cons})). Therefore, in the stage of expansion of the Universe in which the creation of radiation results from the scalar field's dissipation under the significant effects of the boundary contribution, the overall dynamics strongly depends on the nature of the scalar field potential, and on the model parameters.

\subsection{Model II: Creation of radiation driven solely by the scalar field}

 As a second example of warm inflationary model in the presence of scalar and boundary fields, we consider the case in which in the general energy balance equation the scalar fields $\phi$ and $\psi$ decouple in the energy balance equation, and the radiation creation process is fully determined by the scalar field. Hence, radiation creation and the boundary scalar field evolution equations take the form
\begin{eqnarray}
	&&8\pi G\left[\dot{\rho}_m+3H\left(\rho_m+\frac{p_m}{c^2}\right)\right]=\beta \frac{\dot{\phi}^3}{2}e^{\beta \phi},\nonumber \\
	&&\frac{\alpha ^2}{2}\frac{d}{dt}\left(\ddot{\psi}+3H\dot{\psi}+\dot{\psi}^2\right)+\frac{3\alpha ^2}{2}H\dot{\psi}^2=0.
\end{eqnarray}

Particle creation in form of radiation takes place if the condition $\beta \frac{\dot{\phi}^3}{2}e^{\beta \phi}>0$ is satisfied by the model parameter $\beta$.

\paragraph{Dimensionless form of the evolution equations}
In this case, the fundamental equations that describe the warm inflationary model in their dimensionless form are
	\begin{equation}\label{M5}
		\frac{d^2\phi}{d\tau ^2}+3h\frac{d\phi}{d\tau}+v'(\phi)=-\beta \left(\frac{d\phi}{d\tau}\right)^2- \beta v(\phi),
	\end{equation}
	\begin{equation}\label{M6}
		\frac{d}{d\tau}\left[\frac{d^2\psi}{d\tau ^2}+3h\frac{d\psi}{d\tau}+\left(\frac{d\psi}{d\tau}\right)^2\right]+3h\left(\frac{d\psi}{d\tau}\right)^2=0,
	\end{equation}
	\begin{eqnarray}\label{M7}
		2\frac{dh}{d\tau}+3h^2&=&-r_m-e^{\beta \phi}\left[\frac{1}{2}\left(\frac{d\phi}{d\tau}\right)^2-v(\phi)\right]\nonumber\\
		&&+\frac{\alpha ^2}{2}\left(\frac{d^2\psi}{d\tau ^2}+3h\frac{d\psi}{d\tau}\right),
	\end{eqnarray}
	\begin{equation}\label{M8}
		\frac{dr_m}{d\tau}+4hr_m=\frac{\beta }{6}\left(\frac{d\phi}{d\tau}\right)^3 e^{\beta \phi}.
	\end{equation}
	
	Consequently, the relation which provides initial constraints for the system of equations previously presented takes the following form,
	\begin{eqnarray}
		1&-&\frac{e^{\beta \phi(0)}}{3}\left(\frac{\phi'(0)^2}{2}+v(\phi(0))\right)\notag \\
		&=&\frac{\alpha^2}{6}\left(\psi''(0)+3\psi'(0)+\psi'(0)^2\right). \label{c2}
	\end{eqnarray}

\paragraph{Numerical results.} The set of equations (\ref{M5})-(\ref{M8}) were numerically solved by using the following values, $h(0) = 1,\; r_{m}(0) = 0,\; \phi(0) = 2.5,\; \phi'(0) = -2.5,\; \psi(0) = 2,\; \psi'(0) = 1,\; \psi''(0)=1.5$. The parameters were selected to satisfy the condition (\ref{c2}), $\alpha = 0.01$, $\beta_1 = -0.0164$, $\beta_2 = -0.0434$, $\beta_3 =
	-0.0691$, $\beta_4 = -0.0229$. The selection of different values for $\beta$ corresponds to the various forms of the scalar potential, specifically $v_1=0,\; v_2=0.03\phi ^{2},\; v_3=0.008\phi ^{2}+0.01\phi ^{4}$ and $v_4=0.05e^{-0.01\phi }$, respectively.

\begin{figure*}[htbp]
\includegraphics[scale=0.9]{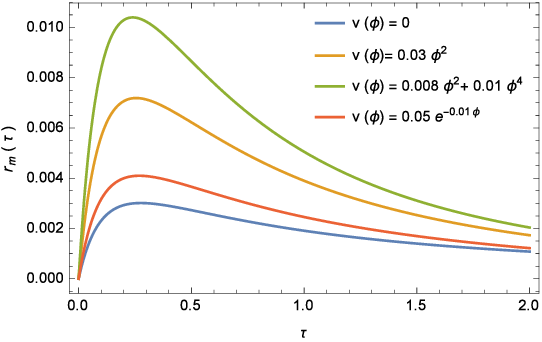}
\includegraphics[scale=0.9]{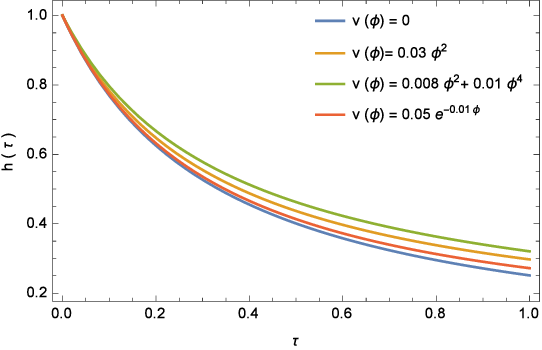}
\caption{Variations as functions of the dimensionless cosmological time of the radiation density $r_m$ (left panel) and of the dimensionless Hubble function $h$ (right panel) for the four considered forms of the scalar field potential in the warm inflationary scenario with scalar field $\phi$ only driven radiation creation.}\label{fig3}
\end{figure*}

 The time variation of the radiation density $r_m$, and of the Hubble function $h$ within the context of the second model warm inflationary model  are illustrated in Fig.~ (\ref{fig3}). 	The function $r_m$ increases due to the decay of the scalar field, reaching a maximum value determined by the form of the scalar potential. Subsequently, $r_m$ decreases as $\phi$ decays at a substantial rate. Notably, the amount of radiation existing after the inflationary era lies below 0.002 for all the chosen scalar potentials. The maximum values of the radiation energy density are obtained for the case of the Higgs type potential. The maximum values of the dimensionless radiation energy $r_m$ for the second warm inflationary model have approximately the same values as in the first inflationary model, occur for the same $\tau \approx 0.247$, and they are presented in Table~\ref{table2}.

\begin{table}[h!]
\begin{center}
\begin{tabular}{|c|c|c|}
\hline
\hline
$v(\phi )$ & $\tau _{\max }$ & $r_{m}^{(\max )}$ \\
\hline
$0$ & $0.253$ & $2.99\times 10^{-3}$ \\
\hline
$0.03\phi ^{2}$ & $0.243$ & $7.15\times 10^{-3}$ \\
\hline
$0.008\phi ^{2}+0.01\phi ^{4}$ & $0.233$ & $10.38\times 10^{-3}$ \\
\hline
$0.05e^{-0.01\phi }$ & $0.258$ & $4.14\times 10^{-3}$\\
\hline
\hline
\end{tabular}
\caption{Maximum values of the radiation energy density for different scalar field potentials in the scalar field only driven radiation creation warm inflationary model.}\label{table2}
\end{center}
\end{table}
	
 The Hubble function monotonically decreases during the warm inflationary phase in the presence of radiation production, and the evolution of the Universe depends on the form of the scalar field potential.
	
\begin{figure*}[htbp]
\includegraphics[scale=0.9]{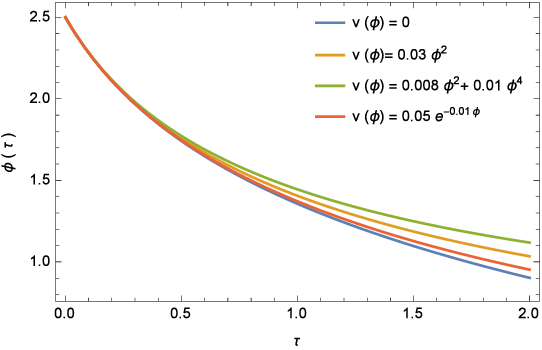}
\includegraphics[scale=0.9]{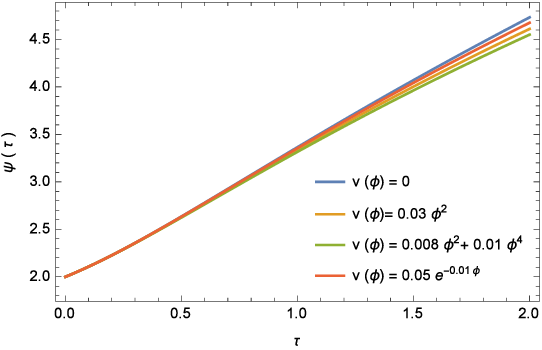}
\caption{Variations as functions of the dimensionless cosmological time of the scalar field  $\phi $ (left panel) and of the boundary contribution  $\psi$ (right panel) for the four considered forms of the scalar field potential in the warm inflationary model with radiation creation due to the scalar field $\phi$ only. }\label{fig4}
\end{figure*}

The variations of the scalar field $\phi$ and of the boundary field $\psi$ are represented, for the second warm inflationary model, in Fig.~(\ref{fig4}). The decay of the scalar field $\phi$ is strongly dependent, at large times, on the form of the potential.  The function $\psi$ increases monotonically with time, showing that the Weyl boundary effects may be present even after the end of the warm inflationary epoch.
	
\subsection{Model III: Emergence of the radiation component from the boundary Weyl vector}

Finally, we consider a third inflationary model, obtained by considering the splitting of the general energy balance equation into the following relations
	
\begin{eqnarray}
	&&8\pi G\left[\dot{\rho}_m+3H\left(\rho_m+\frac{p_m}{c^2}\right)\right]=\alpha ^2\frac{d}{dt}\left(\ddot{\psi}+3H\dot{\psi}+\dot{\psi}^2\right) \nonumber \\
	&&\beta \frac{\dot{\phi}^3}{2}e^{\beta \phi}=\frac{3\alpha ^2}{2}\frac{d}{dt}\left(\ddot{\psi}+3H\dot{\psi}+\dot{\psi}^2\right)+\frac{3\alpha ^2}{2}H\dot{\psi}^2.
\end{eqnarray}

In this model the radiation creation during warm inflation is determined by the boundary Weyl vector $\psi$ only.

\paragraph{Dimensionless form of the warm inflationary evolution equations.}
Therefore, the system of equations that describe the evolution of the early Universe in which the creation of a radiation fluid  is driven only by the boundary Weyl vector is given by
\begin{equation}\label{M9}
	\frac{d^2\phi}{d\tau ^2}+3h\frac{d\phi}{d\tau}+v'(\phi)=-\beta \left(\frac{d\phi}{d\tau}\right)^2-\beta v(\phi),
\end{equation}
\begin{eqnarray}\label{M10}
	&&\frac{\beta }{6}\left(\frac{d\phi}{d\tau}\right)^3 e^{\beta \phi}=\frac{\alpha^2}{2}h\left(\frac{d\psi}{d\tau}\right)^2 \notag  \\
	&&\qquad \qquad+\frac{\alpha^2}{2}\frac{d}{d\tau}\left(\frac{d^2\psi}{d\tau ^2}+3h\frac{d\psi}{d\tau}+\left(\frac{d\psi}{d\tau}\right)^2\right),
\end{eqnarray}
\begin{eqnarray}\label{M11}
	2\frac{dh}{d\tau}+3h^2&=&-r_m-e^{\beta \phi}\left[\frac{1}{2}\left(\frac{d\phi}{d\tau}\right)^2-v(\phi)\right]\nonumber\\
	&&+\frac{\alpha ^2}{2}\left(\frac{d^2\psi}{d\tau ^2}+3h\frac{d\psi}{d\tau}\right),
\end{eqnarray}
\begin{equation}\label{M12}
	\frac{dr_m}{d\tau}+4hr_m=\frac{\alpha^2}{3}\frac{d}{d\tau}\left(\frac{d^2\psi}{d\tau ^2}+3h\frac{d\psi}{d\tau}+\left(\frac{d\psi}{d\tau}\right)^2\right).
\end{equation}

\paragraph{Numerical results.}
The system given by Eqs.~(\ref{M9})-(\ref{M12}) is solved numerically by imposing the following initial conditions: $h(0) = 1,\; r_{m}(0) = 0,\; \phi(0) = 3,\; \phi'(0) = -2.5,\; \psi(0) = 1,\; \psi'(0) = 0.001,\; \psi''(0)=0.01$. The model parameters also satisfy the condition (\ref{c2}), and their values are $\alpha = 1.5,\; \beta_1 = -0.0152,\; \beta_2 = -0.0247, \beta_3 = -0.0592$ and $ \beta_4 = -0.0364$, respectively. The values of the model parameter $\beta$ correspond to the various adopted forms of the scalar field potential $v(\phi)$, elected as $v_1=0, v_2=0.01\phi ^{2},\; v_3=0.04\phi ^{2}+0.001\phi ^{4}$ and $v_4=0.2e^{-0.01\phi }$.

\begin{figure*}[htbp]
\includegraphics[scale=0.9]{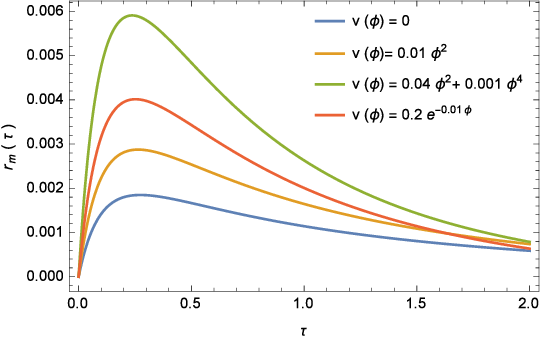}
\includegraphics[scale=0.9]{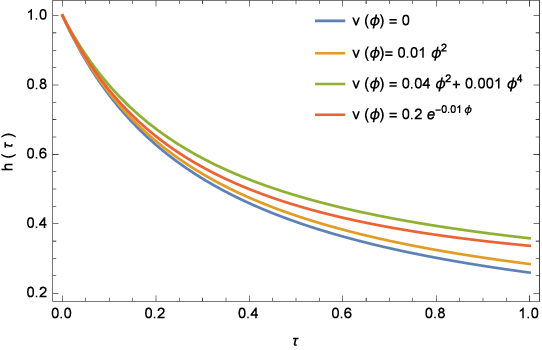}
\caption{Variations as functions of the dimensionless cosmological time of the radiation density $r_m$ (left panel) and of the dimensionless Hubble function $h$ (right panel) for the four considered forms of the scalar field potential in the warm inflationary scenario with the radiation creation driven by the boundary Weyl vector $\psi$.}\label{fig5}
\end{figure*}

	Fig.~(\ref{fig5}) displays the time variation of the radiation energy density $r_m$, and of the Hubble function $h$ within the framework of the third model. The radiation energy density has a similar behavior as in the case of the previous two models, increasing monotonically in the first stages of the cosmological expansion, reaching a maximum, and then, once the rate of the expansion of the Universe takes over the cosmological dynamics, decreasing in time. The specific process of the dynamical creation of radiation depends on the scalar field potential, different potentials giving different maximum values of the density of the radiation fluid.

The maximum values of $r_m$ corresponding to the considered forms of the scalar field potentials are given in Table~\ref{table3}. Generally, the range of numerical values for $r_m^{(max)}$ are similar to those obtained in the previously considered warm inflationary models.  It is important to note that despite the primary influence of the dynamics of the Weyl vector on particle creation, the rate of radiation generation remains consistent with that of the first two models.  The time interval for the maximum values is $\tau \approx 0.251$. Moreover, the energy density of the radiation sector also tends asymptotically to the value 0.001, the same as for the other two models.

\begin{table}[h!]
\begin{center}
\begin{tabular}{|c|c|c|}
\hline
\hline
$v(\phi )$ & $\tau _{\max }$ & $r_{m}^{(\max )}\times 10^{3}$ \\
\hline
$0$ & $0.265$ & $1.85$ \\
\hline
$0.01\phi ^{2}$ & $0.254$ & $2.88$ \\
\hline
$0.04\phi ^{2}+0.001\phi ^{4}$ & $0.237$ & $5.89$ \\
\hline
$0.2e^{-0.01\phi }$ & $0.248$ & $3.99$\\
\hline
\hline
\end{tabular}
\caption{Maximum values of the radiation energy density in the warm inflationary model for different scalar field potentials, in the warm inflationary model with the matter source term determined by the boundary Weyl vector only.}\label{table3}
\end{center}
\end{table}

The Hubble function decreases in time, indicating an expansionary evolution during matter creation.

\begin{figure*}[htbp]
\includegraphics[scale=0.9]{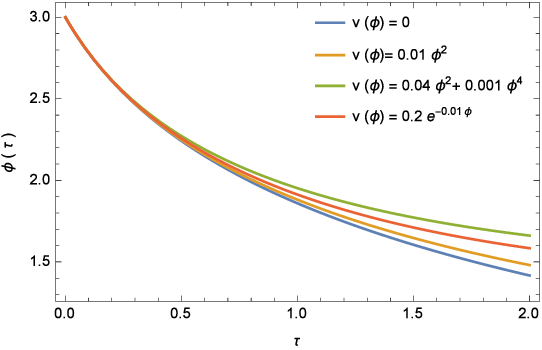}
\includegraphics[scale=0.9]{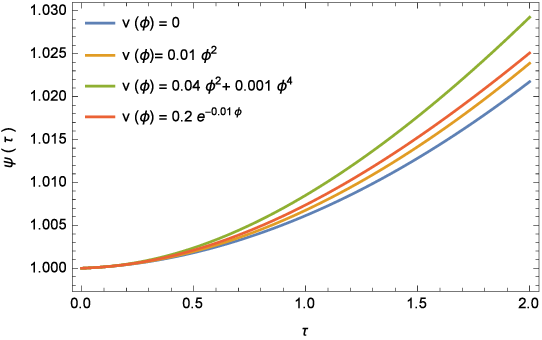}
\caption{Variations as functions of the dimensionless cosmological time of the scalar field  $\phi $ (left panel) and of the boundary contribution  $\psi$ (right panel) for the four considered forms of the scalar field potential in the warm inflationary model with radiation creation determined by the boundary Weyl vector only. }\label{fig6}
\end{figure*}

The two scalar field's dynamics are portrayed in Fig.~(\ref{fig6}), where $\psi$ determines the radiation creation rate, while $\phi$ directly contributes to the cosmological dynamics. The scalar field transfers its energy to the surrounding environment, which leads to the dissipation of its energy, and to the decrease of its magnitude in time. The boundary Weyl field contribution shows a rapid rate of change as it interacts with the radiation component.

\section{Discussions and final remarks} \label{conclusion}

In the present paper we have considered an extension of the gravitational action of standard general relativity by considering the contribution of the boundary terms to the general dynamics, a contribution whose dynamical effects are generally neglected in the study of gravitational phenomena. Following the studies initiated in \cite{BR1,BR2,B3}, we have assumed that the background geometry in which the boundary terms may play an important role is of Weyl type. The variation of the connection is generally expressed in terms of the covariant derivative $\nabla _\mu$ of the variation of the metric tensor. By substituting $\nabla _\mu$ with its Weylian counterpart, and by extensively using the formalism of Weyl geometry, one can obtain the expression of the boundary terms as given by Eq.~(\ref{40}). This boundary term, depending on the Weyl vector, and on its covariant derivative, gives a non-zero contribution to the field equations. We have also implicitly assumed that the Hilbert-Einstein type action of the gravitational sector can be related to the Weyl geometry, and it emerges as the result of the symmetry breaking of a conformally invariant Weyl type action, as initially introduced by Weyl \cite{Weyl}, and later on extensively discussed in \cite{Gh1,Gh2,Gh3,Gh4,Gh5, Gh6}. By introducing an auxiliary scalar field $\phi$, the quadratic term in the Weyl scalar, $\tilde{R}^2$, can be linearized. Then, by means of a Stueckelberg type mechanism, in which the scalar field is substituted by its vacuum expectation value, the gravitational sector decouples from the scalar field one, and thus the Hilbert-Einstein Lagrangian density is recovered. Scalar fields, of geometric origin, can be generated in the integrable Weyl geometry, in which the Weyl vector is the gradient of a scalar function, either by using the linearization procedure, or by adding the fields by hand. In our approach,  the symmetry of the Weylian Einstein tensor requires that the strengths $F_{\mu \nu}$ of the Weyl vector vanishes, which leads  to the integrable Weyl geometry, in which a scalar field can be naturally generated from the Weyl vector.

However, in our approach, and for the sake of generality, we also consider the possibility of the existence of a scalar field of a physical (non-geometric) origin. Once we include the scalar field, the general system of field equations in the presence of boundary terms can be formulated in the form (\ref{fe}), and it represent a physical model in which, along the matter term, two distinct scalar fields contribute to the gravitational dynamics: the standard scalar field $\phi$, and a second scalar field $\psi$, describing the contribution of the Weyl type boundary terms. Even that these two fields may be identical, in the present approach we have assumed that they are distinct in their physical nature. This assumption is valid if the scalar field $\phi$  is of a non-geometrical, physical nature.

It has already been mentioned in \cite{Gh4} that in the canonical Weyl geometric Lagrangian, the  coupling of the Weyl vector $\omega_\mu$ to the neutral Higgs boson of the Standard Model of the elementary particles $\sigma$ is given by
\bea
L_H=\frac18\,\sqrt{-g}\, \alpha ^2\,\omega _\mu \omega ^\mu\,\sigma^2+  O(\sigma^2/M_p^2).
\eea
The above Lagrangian represents the only direct coupling of the Standard Model particles to the Weyl gauge boson. In the very early Universe, assuming there was no Higgs boson, this coupling can generate the Higgs particle from the Weyl vector boson fusion
\bea
\omega_\mu+\omega_\mu\rightarrow \sigma+\sigma.
\eea

Since  $\omega_\mu$ originates from the non-metric Weyl geometry,   the Higgs boson  itself is generated from geometry. Since the Higgs boson determines the masses of the Standard Model particles,  it follows  that all the masses of the theory  originate from $\omega_\mu$ and Weyl geometry \cite{Gh4}, without the requirement of the existence of any additional degrees of freedom \cite{Gh4}.

It is also important to point out  that what from the Weyl geometry perspective looks like matter creation from the Weyl vector  $\omega _\mu$, from a Riemannian viewpoint obtained after the symmetry breaking,  $\omega_\mu$ corresponds to a scalar field of the basic theory.

As an application of the gravitational field equations in the presence of the Weylian boundary terms we have considered, within the framework of the warm inflationary scenario, the creation of radiation during the very early expansionary phase of the Universe. In our model, the cosmological dynamics is determined by two scalar fields $\psi$ and $\phi$, one originating from the boundary of the Universe, while the second is a scalar field having its origin either in the Weyl geometry, or from other physical processes. In order to have a consistent theoretical approach, we have introduced the dissipative scalar field by including it in the total action, by following the approach of \cite{dis}. Thus, the energy momentum tensor can be obtained in systematic and rigorous way from a variational principle.

The starting point in the analysis of the radiation  generation is the generalized energy balance equation, formulated in the FLRW geometry, which can be split in several ways to describe radiation creation. We have considered in detail three distinct warm inflationary models, depending on the source term for radiation. In the first model the source term in the energy balance equation for radiation contains both the contributions of the scalar field $\phi$, and of the boundary field $\psi$. In the second model, the radiation  source term is determined only by the scalar field $\phi$, while in the third model the source term is determined by the boundary field $\psi$. In all three cases we have investigated in detail the cosmological evolution, which shows a common pattern. The radiation energy density increases from a zero value to a maximum, and then decreases, once the particle creation rate falls below the Hubble expansion rate. The particle creation phase takes place during the expansionary phase of the Universe, and its quantitative description depends on the form of the scalar field potential. The scalar field decreases during the radiation creation due to its energy transfer to radiation. Interestingly enough, the boundary scalar field $\psi$ increases during the early Universe evolution. Its presence in the late Universe may play an important role in the cosmological dynamics, acting, for example, as an effective cosmological constant, triggering the recent accelerated expansion of the Universe \cite{BR1,BR2,B3}.

An important physical and cosmological quantity is the maximum temperature reached by the radiation fluid during warm inflation. A relation between the radiation energy density and the temperature of the Universe can be established by means of the Stefan-Boltzmann law. Therefore, once the evolution of the density of the radiation fluid is known, the temperature of the Universe can be obtained through Eq.~(\ref{SB}), as
\begin{equation}
	T(\tau)=\left(\frac{3H_0^2}{8\pi G\sigma}\right)^{1/4}r_m^{1/4}(\tau),
\end{equation}
where $r _m=\frac{8 \pi G}{3H_0^2}\rho_m$ is the dimensionless form of the energy density for the radiation fluid.
We assume for the early Universe a maximum radiation  temperature of the order of $T_{max}=10^{16}\; \rm{GeV}$, which also corresponds to the maximum temperature of the reheating in the post-inflationary regime. Hence, the value of the Hubble function at the instant after the beginning of the expansion of the Universe can be obtained from the relation
\begin{equation}
	T_{max}=2.147\times 10^{10}\times H_0^{1/2}\times \left(r_{m_{max}}\right)^{1/4}=10^{16}\; \rm GeV.
\end{equation}

From the analysis of the three considered models, we find the maximum values of the energy density within the range of $r_{m_{max}}\in [0.00185, 0.01038]$. Accordingly, the value of the Hubble constant in the early Universe varies, in our models,  within the interval $H_0 \in \left[2.866\times 10^{38} , 6.789\times 10^{38} \right]\; \rm s^{-1}$. Note that the present day value of the Hubble constant is of the order of $H_0=2.2\times 10^{-18}\;{\rm s}^{-1}$. An important characteristic of warm inflation is that the radiation fluid temperature must be larger than the Hubble function, $T>H$. This condition is related to the requirement that thermal fluctuations must be larger than quantum fluctuations, and thus they are at the origin of the formation of the large scale structure in the Universe.

An important problem is the comparison of the predictions of the present warm inflationary model with observations. To investigate the observational constraints on the warm inflationary one introduces the first, and the second slow-roll parameters, defined according to \cite{W48,W49}
\be\label{slowroll1}
\epsilon _1 =q+1=  - \frac{\dot {H}}{H^2}=- \frac{3}{2h^2}\frac{dh}{d\tau },
\ee
and
\bea\label{slowroll2}
\hspace{-0.9cm}\epsilon_2=\frac{\dot{\epsilon} _1}{H\epsilon _1}
\equiv  \frac{\ddot{H}}{H\dot{H}}-\frac{2\dot{H}}{H^2}=\frac{3}{2 h^2}\frac{d^2h/d\tau ^2}{d h/d\tau } - \frac{3}{ h^2}\frac{d h}{d\tau },
\eea
respectively.

To measure the amount of cosmic expansion during  inflation one uses the number of e-folds $\mathcal{N}$, defined as
\bea\label{efoldtau}
\hspace{-0.3cm}\mathcal{N} &=& \int_{t_\star}^{t_e} H \; dt = \frac{2}{3}\int_{\tau_\star}^{\tau_e} {h} \; d\tau,
\eea
where $\tau_{e}$ is the re-scaled time at the end of inflation, while  $\tau_\star$ is the re-scaled time at the horizon crossing. The amplitude of the scalar perturbations $\mathcal{P}_s$ can be obtained as \cite{W48,W49}
\begin{equation}\label{pswarm}
\mathcal{P}_s = \frac{25}{4} \left( { H \over \dot\phi }\right)^2 \left(\delta\phi \right)^2,\;\;\left(\delta\phi \right)^2=\frac{\sqrt{\Gamma_1 H}T}{2\pi^2},
\end{equation}
while the spectral scalar index is defined as  \cite{W48,W49}
\begin{eqnarray}\label{nsaswarm}
n_s - 1 &=& {d\ln(\mathcal{P}_s) \over d\ln(k)}={d\ln(\mathcal{P}_s) \over d\ln(a)}\frac{d\ln (a) }{d\ln (k)}=\frac{1}{H}\frac{d}{dt}\ln(\mathcal{P}_s)=\nonumber\\
&&\frac{3}{2}\frac{1}{h}\frac{d}{d\tau}\ln \mathcal{P}_{s} (\tau).
\end{eqnarray}

In terms of the function $Q$, defined in Eqs.~(\ref{Q1}) and (\ref{Q2}), the slow-roll parameters can be expressed as
\begin{equation}\label{vsrp}
\epsilon_1 = {M_P^2 \over 2(1+Q)} {V^{\prime 2}(\phi) \over V^2(\phi)}\;, \qquad \epsilon_2 = {\dot{\epsilon}_1 \over H \epsilon_1}\;.
\end{equation}

One can also introduce two other slow-roll parameters, defined in terms of the potential and of $Q$ as,
\begin{equation}\label{etabeta}
\eta = {M_P^2 \over (1+Q)} \; {V''(\phi) \over V(\phi)},\; \beta = {M_P^2 \over (1+Q)} \; {V'(\phi) \Gamma'(\phi) \over V(\phi) \Gamma(\phi)}\;.
\end{equation}

 $\epsilon_2$ can be also expressed  as
\begin{equation}
  \epsilon_2 = -2 \eta + 4 \epsilon_1 + {Q \over (1+Q)} \; \left( \beta - \epsilon_1 \right)\;.
\end{equation}

These parameters can be estimated for the present warm inflationary models, and a comparison with the observational data can be performed to test the viability of the models. This comparison will be presented in a future work.

Radiation generation has very important effects on the evolution and dynamics of the very early Universe, as well as for the present day structure and composition of the cosmic environment. In the present approach this process depends on the scalar field - Weyl boundary vector - photon fluid interaction mechanism, and
on the parameters of the particle physics models necessary to describe the physical properties of the newly created particles.  However,  presently there is no definite theory giving the numerical values of the physical parameters describing, for example,  the interaction between a scalar field and a radiation fluid. In the present work we have proposed an essentially  geometric approach to the description of the radiation creation processes from two scalar fields, both having a geometric origin. The present results may represent some basic theoretical tools necessary for the in-depth understanding of the complex physical processes that took place in the early Universe.

\appendix

\section{The variation of Ricci tensor}\label{appa}

The Riemannian curvature tensor is defined in terms of the connection as
\begin{equation}
	R^{\lambda}_{\mu \nu \sigma} = \partial_{\nu}\Gamma^{\lambda}_{\mu \sigma}-\partial_{\sigma}\Gamma^{\lambda}_{\mu \nu}+\Gamma^{\lambda}_{\nu \rho}\Gamma^{\rho}_{\mu \sigma}-\Gamma^{\lambda}_{\sigma \rho}\Gamma^{\rho}_{\mu \nu},
\end{equation}
and its variation takes the following form,
\begin{eqnarray}
	\delta R^{\lambda}_{\mu \nu \sigma} = \partial_{\nu}\delta\Gamma^{\lambda}_{\mu \sigma}&-&\partial_{\sigma}\delta \Gamma^{\lambda}_{\mu \nu}+\delta\Gamma^{\lambda}_{\nu \rho}\Gamma^{\rho}_{\mu \sigma}+\Gamma^{\lambda}_{\nu \rho}\delta\Gamma^{\rho}_{\mu \sigma} \nonumber \\
	&-&\delta\Gamma^{\lambda}_{\sigma \rho}\Gamma^{\rho}_{\mu \nu}-\Gamma^{\lambda}_{\sigma \rho}\delta\Gamma^{\rho}_{\mu \nu}.
\end{eqnarray}

The covariant derivative of the Christoffel symbol is given by
\begin{equation}
	\nabla_{\nu}\delta\Gamma^{\lambda}_{\mu \sigma}=\partial_{\nu}\delta\Gamma^{\lambda}_{\mu \sigma}+\Gamma^{\lambda}_{\nu \rho}\delta\Gamma^{\rho}_{\mu \sigma}-\Gamma^{\rho}_{\nu \mu}\delta\Gamma^{\lambda}_{\rho \sigma}-\Gamma^{\rho}_{\nu \sigma}\delta\Gamma^{\lambda}_{\mu \rho},
\end{equation}
therefore variation of the Riemannian curvature tensor can be obtained as
\begin{equation}
	\delta R^{\lambda}_{\mu\nu\sigma}=\nabla_{\nu}\delta\Gamma^{\lambda}_{\mu \sigma}-\nabla_{\sigma}\delta\Gamma^{\lambda}_{\mu \nu}.
\end{equation}

Thus, the variation of the Ricci tensor can be easily obtained, considering $\delta R_{\mu\nu}=\delta R^{\lambda}_{\mu\lambda\sigma}$,
\begin{equation}
	\delta R_{\mu\nu}=\nabla_{\lambda}\delta\Gamma^{\lambda}_{\mu \sigma}-\nabla_{\sigma}\delta\Gamma^{\lambda}_{ \mu \lambda} \equiv \nabla _{\lambda}\delta \Gamma _{\mu \nu}^{\lambda}-\nabla _{\nu }\delta \Gamma _{\mu \sigma}^{\sigma }.
\end{equation}

\section{The variation of Christoffel symbols}\label{appb}

The Christoffel symbols are defined as
\begin{equation}
	\Gamma^{\lambda}_{\mu \nu}=\frac{1}{2}g^{\lambda \sigma}\left(\partial_{\mu}g_{\sigma \nu }+\partial_{\nu}g_{\mu \sigma}-\partial_{\sigma}g_{\mu \nu}\right).
\end{equation}

Their variation is given by
\begin{eqnarray}
	\delta\Gamma^{\lambda}_{\mu \nu}&=&\frac{1}{2}\delta g^{\lambda \sigma}\left(\partial_{\mu}g_{\sigma \nu }+\partial_{\nu}g_{\mu \sigma}-\partial_{\sigma}g_{\mu \nu}\right) \notag \\
	&+&\frac{1}{2}g^{\lambda \sigma}\left(\partial_{\mu}\delta g_{\sigma \nu }+\partial_{\nu}\delta g_{\mu \sigma}-\partial_{\sigma}\delta g_{\mu \nu}\right),
\end{eqnarray}
and by using the relation $\delta g_{\mu \lambda }=-g_{\mu \nu }g_{\lambda \sigma }\delta g^{\nu \sigma}$, we find successively
\begin{eqnarray}
	\delta\Gamma^{\lambda}_{\mu \nu}&=&-\frac{1}{2}\delta g^{\lambda \rho}g^{\sigma \epsilon}\delta g_{\rho \epsilon}\left(\partial_{\mu}g_{\sigma \nu }+\partial_{\nu}g_{\mu \sigma}-\partial_{\sigma}g_{\mu \nu}\right) \notag \\
	&+&\frac{1}{2}g^{\lambda \sigma}\left(\partial_{\mu}\delta g_{\sigma \nu }+\partial_{\nu}\delta g_{\mu \sigma}-\partial_{\sigma}\delta g_{\mu \nu}\right) \\
	&=&- g^{\lambda \sigma}\delta g_{\sigma \epsilon}\Gamma^{\epsilon}_{\mu \nu}+\frac{1}{2}g^{\lambda \sigma}\left(\partial_{\mu}\delta g_{\sigma \nu }+\partial_{\nu}\delta g_{\mu \sigma}-\partial_{\sigma}\delta g_{\mu \nu}\right) \notag \\
	&=& \frac{1}{2}g^{\lambda \sigma}\left(\partial_{\mu}\delta g_{\sigma \nu }+\partial_{\nu}\delta g_{\mu \sigma}-\partial_{\sigma}\delta g_{\mu \nu}-2\delta g_{\sigma \epsilon}\Gamma^{\epsilon}_{\mu \nu}\right). \notag
\end{eqnarray}

Therefore, the following relation can be obtained
\begin{align}
	\delta\Gamma^{\lambda}_{\mu \nu}&=\frac{1}{2}g^{\lambda \sigma}\left(\partial_{\mu}\delta g_{\sigma \nu}-\Gamma^{\epsilon}_{\mu \nu}\delta g_{\epsilon\sigma}-\Gamma^{\epsilon}_{\mu \sigma}\delta g_{\epsilon\nu}\right. \notag \\
	&\qquad + \partial_{\nu}\delta g_{\mu \sigma}-\Gamma^{\epsilon}_{\nu \sigma}\delta g_{\epsilon\mu}-\Gamma^{\epsilon}_{\nu \mu}\delta g_{\epsilon\sigma} \notag \\
	&\qquad \left. - \partial_{\sigma}\delta g_{\mu \nu}+\Gamma^{\epsilon}_{ \sigma\mu}\delta g_{\epsilon\nu}+\Gamma^{\epsilon}_{\sigma\nu}\delta g_{\epsilon\mu} \right),
\end{align}
and consequently,
\begin{equation}
	\delta\Gamma^{\lambda}_{\mu \nu}=\frac{1}{2}g^{\lambda \sigma}\left(\nabla_{\mu}\delta g_{\sigma \nu}+\nabla_{\nu} \delta g_{\mu\sigma}-\nabla_{\sigma}\delta g_{\mu \nu} \right).
\end{equation}

\end{document}